\documentclass[twocolumn,superscriptaddress]{revtex4-1}
\pdfoutput=1

\usepackage{amssymb}
\usepackage{amsmath}
\usepackage{graphicx}

\begin{document}

\title{The conformation of thermoresponsive polymer brushes probed by optical reflectivity}

\author{Siddhartha Varma}
\affiliation{Univ. Grenoble Alpes, LIPHY, F-38000 Grenoble, France}
\affiliation{CNRS, LIPHY, F-38000 Grenoble, France}
\author{Lionel Bureau}\email{lionel.bureau@ujf-grenoble.fr}
\affiliation{Univ. Grenoble Alpes, LIPHY, F-38000 Grenoble, France}
\affiliation{CNRS, LIPHY, F-38000 Grenoble, France}
\author{Delphine D{\'e}barre}
\affiliation{Univ. Grenoble Alpes, LIPHY, F-38000 Grenoble, France}
\affiliation{CNRS, LIPHY, F-38000 Grenoble, France}

\begin{abstract}

{\small We describe a microscope-based optical setup that allows us to perform space- and time-resolved measurements of the spectral reflectance of transparent substrates coated with ultrathin films. This technique is applied to investigate the behavior in water of thermosensitive polymer brushes made of poly(N-isopropylacrylamide) grafted on glass. We show that spectral reflectance measurements yield quantitative information about the conformation and axial structure of the brushes as a function of temperature. We study how molecular parameters (grafting density, chain length) affect the hydration state of a brush, and provide one of the few experimental evidence for the occurrence of vertical phase separation in the vicinity of the lower critical solution temperature of the polymer. The origin of the hysteretic behavior of poly(N-isopropylacrylamide) brushes upon cycling the temperature is also clarified. We thus demonstrate that our optical technique allows for in-depth characterization of stimuli-responsive polymer layers, which is crucial for the rational design of smart polymer coatings in actuation, gating or sensing applications.}

\end{abstract}

\maketitle

\section{Introduction}

The use of polymer brushes for the design of advanced functional substrates has gained a rapidly growing interest over the past fifteen years \cite{revBrush1,revBrush2}. These ultrathin films, composed of macromolecules tethered by one end to a substrate, can be bound to a variety of surfaces with tight control over the chemical composition of the macromolecules, their architecture, length and surface density \cite{revBrushAppli2}. Particular attention has been paid to the elaboration of brushes made of stimuli-responsive polymers, in order to design  surfaces whose properties can be altered on demand via the control of an external parameter \cite{revBrushAppli1} (pH, temperature, light, ionic strength...). Such smart substrates exploit the fact that, upon application of the external stimulus, the grafted macromolecules undergo a change in conformation, which in turn affects the properties of the functionalized surface\cite{revResponsiveBrush}. This has given rise to a breadth of applications ranging from the reversible switching of wetting, to the design of nanofluidic valves or chemical gates\cite{revBrush1,revBrushAppli1}. It has also fostered research focusing on the use of stimuli-responsive polymer brushes to control the interactions between a surface and biological objects\cite{revBio1,revBio3}. For instance, since the seminal work of Okano {\it et al.} \cite{cellsheet} that reported the controlled detachment of mammalian cells cultured on thermo-responsive polymer layers, smart polymer brushes have been used in a number of biology-relevant applications \cite{revBio1,revBio3} such as protein immobilization/release from surfaces\cite{protScience}, tissue engineering\cite{tissue} and selective chromatography\cite{chromato}. 

These applications rely on the fact that adsorption of proteins on a brush-grafted surface depends strongly on the conformation of the tethered chains \cite{Avi2,LeckbandProt1}, and can thus be switched on or off through appropriate variations of the external stimulus. In this context, the rational design of stimuli-responsive polymer brushes requires the central questions to be addressed, including how parameters of the brush ({\it e.g.} its grafting density) influence: the conformation of the grafted chains, the overall structure of the brush, its response to the applied stimulus, and {\it in fine} its usage properties for the target application \cite{Avi3}.

Among the number of polymers that have been chosen or designed in the above-mentioned works, poly(N-isopropylacrylamide) (PNIPAM) is one of the most widely employed. It is a thermoresponsive polymer that exhibits, in pure water, a Lower Critical Solution Temperature (LCST) at about 32$^{\circ}$C\cite{Wu2}. Studies performed on dilute solutions of PNIPAM in water show that, as the solution temperature is increased across the LCST, individual  polymer chains switch from a well-solvated, swollen coil configuration to a collapsed, globular conformation \cite{Wu2}. This reversible variation in solubility and the associated change in chain dimension and hydrophilicity occurs at a LCST close to physiological temperatures, making PNIPAM a choice polymer for surface functionalization in biological applications. 

So far, the conformation of PNIPAM brushes have been investigated experimentally using well-established techniques that are sensitive to the presence of thin surface-bound layers: 

(i) Surface Plasmon Resonance\cite{SPR} (SPR)

(ii) Quartz-Crystal Microbalance\cite{QCM1,QCM2,QCM3,QCM4,AFM-QCM1,AFM-QCM2} (QCM)

(iii) force measurement techniques such as the Surface Forces Apparatus\cite{LeckbandSFA,bureau1} (SFA) or Atomic Force Microscopy\cite{AFM1,AFM2,AFM3,AFM-QCM1,AFM-QCM2} (AFM)

(iv) Neutron Reflectivity \cite{Neutron1,Neutron2,Neutron3} (NR)

(v) in-situ Spectral Ellipsometry \cite{collapseEllipso,Ellipso2} (SE).

While SPR and QCM studies give access to the qualitative evolution of the brush refractive index or effective mass with temperature, force measurements allow the inference of the temperature-dependent brush swelling from the range of steric repulsive forces measured, and only NR and SE (or, equivalently, multi-angle ellipsometry\cite{Ellipso3} or scanning angle reflectometry\cite{gast}) studies can give quantitative insights regarding the evolution of the axial structure of brushes, {\it i.e.} of the monomer volume fraction as a function of the sub-micrometer distance from the grafted surface. 

Brought together, the conclusions from these various studies have led to the following qualitative picture: 

(i) grafting density and chain length greatly affect the magnitude of conformational change upon temperature variations \cite{Neutron2,bureau1}, 

(ii) the thermal response of a PNIPAM brush is observed to be hysteretic during a heating/cooling cycle \cite{SPR,QCM4,QCM3}, and 

(iii) densely grafted brushes may collapse non-homogeneously within their thickness and display a so-called vertical phase separation around the bulk LCST \cite{Neutron3,collapseEllipso}.

All these phenomena may directly influence the performances of a brush in a given application (the sharpness of the ``switch'', its efficiency, dynamics and reversibility) and need to be accurately characterized. Moreover, point (iii) above is theoretically predicted to be a direct result of the bulk PNIPAM phase diagram in water (i.e. of the system free energy of mixing), which is still a matter of uncertainty despite years of research and data accumulation \cite{Avi2,Winnik}.

Overall, there is still a clear need, in the field of stimuli-responsive brushes, of experimental techniques that allow combining the quantitative outputs of {\it e.g.} neutron reflectometry with the ease of access and implementation of laboratory-scale setups, in order to quickly and efficiently screen the effect of the characteristics of a brush on its response. Additionally, a larger body of data related to vertical phase separation and hysteresis of PNIPAM brushes would allow clearer identification of the mechanisms at play and opportunities to test theoretical predictions. 

We address these two aspects in the present work. Using an original optical setup based on Reflection Interference Contrast Microscopy (RICM), we measure over the visible range the space- and time-resolved spectral reflectance of PNIPAM brushes elaborated by a ``grafting-from'' procedure. A careful analysis of these spectra allows us to go beyond qualitative studies\cite{refracto}, and to retrieve quantitatively the density profiles of the brushes and study their dependence on temperature. We thus provide a confirmation for the existence of vertical phase separation within PNIPAM brushes, but also extend previous experiments and confrontations with theory by studying how chain length and grafting density affect this phenomenon. Furthermore, we perform an analysis of the density profiles of the swollen brushes that provides us with an unprecedented level of information regarding the length, grafting density and polydispersity of the brushes, making our technique a very powerful tool for {\it in situ} and non-destructive characterization of brushes that overcome many of the limitations encountered with other characterization methods. Finally, the study of the hysteretic response of PNIPAM brushes during a heating-cooling cycle permits for the first time to discriminate between two plausible mechanisms at the origin of hysteresis.


\section{Results and discussion}
\label{sec:res}

To study the influence of the grafting density and chain length on the thermal properties of PNIPAM brushes, we have used a set of 12 brushes grown on glass surfaces (see Supporting Information for details). Brushes were elaborated with 3 different grafting densities, to which we refer to as HD (high), MD (medium) and LD (low) densities. HD brushes with 10 different chain lengths were prepared. The characteristics of the brushes studied are summarized in Table \ref{table1}.
\begin{table}[h!]
  \caption{Characteristics of PNIPAM brushes. $^{(a)}$dry thickness measured by ellipsometry.  $^{(b)}$ swelling ratio determined from reflectivity spectra at 25$^{\circ}$C. $^{(c)}$grafting density estimated from Eq. \ref{eq:sig}, with $\nu=1/2$ and an average swelling ratio of $\alpha=3.1$ for all HD brushes. MD and LD densities estimated from their dry thicknesses compared to HD control wafers. $^{(d)}$number of monomer per chain estimated from Eq. \ref{eq:1}. $^{(e)}$Ratio of the Flory radius ($R_F=aN^{3/5}$) and the estimated average distance between tethering points $d=\sigma^{-1/2}$. $^{(f)}$Ratio of the collapsed radius ($R_c=aN^{1/3}$) and the estimated average distance between tethering points. These two ratios are larger than 1 for all the samples, which shows that our layers are all in the brush regime, independently of their hydration state.}
  \label{table1}
  \begin{tabular}{ccccccc}
    \hline
    Brush&$h_{dry}$$^{(a)}$ & $\alpha$$^{(b)}$ & $\sigma$$^{(c)}$ &  $N$$^{(d)}$ & $\frac{R_F}{d}$$^{(e)}$ & $\frac{R_c}{d}$$^{(f)}$\\
    \# & (nm) & & (nm$^{-2}$) & & \\
    \hline
    HD1 &  200 & 2.8 & 0.3 & 3100 & 41 & 4.8\\
    HD2 & 167 & 3.2 & 0.3 & 2580 & 36 & 4.5\\
    HD3 & 138 & 3.4 & 0.3 & 2130 & 33 & 4.2\\
    HD4 & 108 & 2.75 & 0.3 & 1670 & 28 & 3.9\\
    HD5 & 90& 3 & 0.3 & 1390 & 25 & 3.7\\
    HD6 & 70 & 2.95 & 0.3 & 1080 & 22 & 3.4\\
    HD7 & 62 & 3.2 & 0.3 & 960 & 20 & 3.2\\
    HD8 & 22 & 3.4 & 0.3 & 340 & 11 & 2.3\\
    HD9 & 17 & 2.8 & 0.3 & 260 & 9 & 2.1\\
    HD10 & 13 & 3.2 & 0.3 & 200 & 8 & 1.9\\
    MD & 40  & 6.5 & 0.088 & 2150 & 18 & 2.3\\
    LD & 29 & 11 & 0.045 & 3100 & 16 & 1.9\\
    \hline
  \end{tabular}
\end{table}

\subsection{Brush collapse and vertical phase separation}
\label{subsec:Temp}

\begin{figure}
$$
\includegraphics[width=8cm]{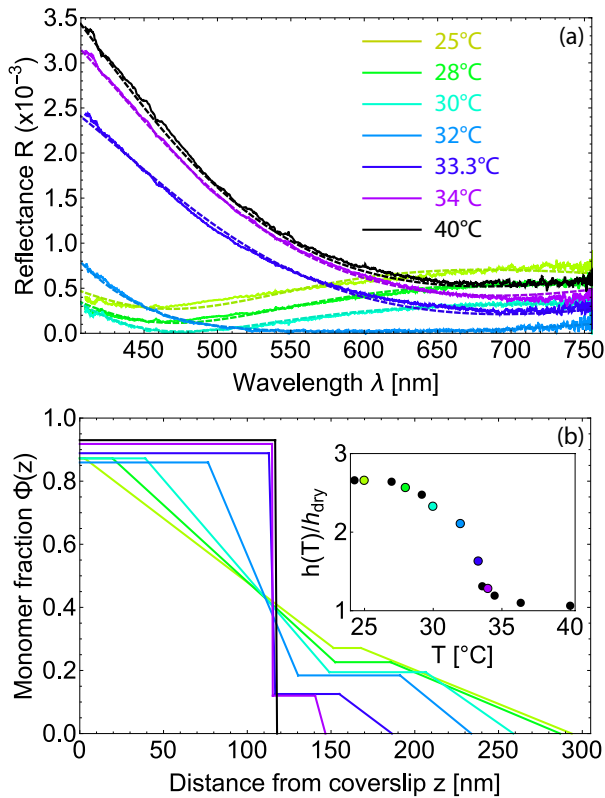}
$$
\caption{\label{fig:spectrawithfits} (a), reflectance spectra from brush HD4 as a function of temperature. Solid lines are the measured spectra, dashed lines are the best fits obtained with the profiles displayed in (b). Inset, swelling ratio as a function of temperature, showing a marked transition around the LCST.}
\end{figure}

Our optical setup is described in details in the Materials and Methods section. We provide in Fig. \ref{fig:spectrawithfits}(a) a set of reflectance spectra measured at different temperatures for one given brush (high grafting density sample HD4): both the magnitude and the wavelength-dependence of the reflectance evolve with temperature, and exhibit marked variations in the vicinity of the bulk LCST of the polymer. By fitting these curves using a simple model described and discussed in the Materials and Methods section, density profiles as a function of the distance from the grafted surface ($\phi(z)$) can thus be extracted from these spectra.

Below and around the LCST, we find that under our modelling assumptions, reflectance curves cannot be properly described by using $\phi(z)$ profiles having a simple steplike shape or composed of a layer of uniform density and a single linear decay (see Fig. \ref{fig:spectra2D} and movie 1). Instead, a satisfactory quantitative description of the reflectance curves could be reached when accounting for more complex density profiles, as illustrated in Fig. \ref{fig:spectrawithfits}(b).

At $T=25^{\circ}$C, a decaying density profile composed essentially of two linear gradients is able to capture the main features of the spectra. 

Upon increasing $T$ from ambient up to about 33$^{\circ}$C, the following features emerge in the profiles: 

(i) a near-wall region of high and uniform  density builds up and extends away from the substrate, 

(ii) $\phi(z)$ adopts a shape indicative of a two-step decay, with a dense inner region topped by a more dilute outer one (such a shape is referred to as a ``two-phase profile" in the following), 

(iii) the graded zone separating the dense and dilute regions becomes thinner, and 

(iv) the distance from the substrate at which $\phi$ drops to zero, {\it i.e.} the brush height, is reduced (see Fig. \ref{fig:spectrawithfits}(b), inset).

At temperatures above 34$^{\circ}$C, we find that experimental data are equally well fitted using a ``single-phase'' profile made of a uniform region of high density surmounted by a single decaying zone, with an overall brush height that approaches the dry thickness as temperature is raised (see movie 1).

Finally, far above the LCST, at $T=40^{\circ}$C, we find that  spectral reflectance curves can be fitted using a simple steplike density profile with a constant volume fraction $\phi=0.9-1$ and a height comparable, to within about 10\%, with the dry thickness of the samples determined independently by ellipsometry. This shows that well above the polymer LCST, PNIPAM brushes adopt the expected dense and collapsed conformation.

Such qualitative features, and in particular the two phases needed in order to properly describe the spectra in the vicinity of the LCST, are found in the $\phi(z)$ profiles of all HD brushes with dry thicknesses $h_{dry}\gtrsim 60$ nm, as illustrated in Fig. \ref{fig:fitsatthetransition} provided in Supporting Information. This shows that the collapse of a brush occurs non-uniformly over its thickness, and that such vertical structuring is not affected by chain length for $N\gtrsim 1000$. For HD brushes of $h_{dry}\lesssim 20$nm, reflectance curves are well fitted, over the whole temperature range, by using a single-phase profile whose extension decreases when $T$ is increased. We cannot discriminate, in the vicinity of the LCST, between single- and two-phase profiles, which yield equally good fits. This is due to the technique reaching its sensitivity limit, as discussed in Supporting Information. Therefore, we cannot conclude on whether vertical structuring is affected by chain length for the thinnest brushes investigated. However it is worth noting that our technique still provides access to a quantitative monitoring of the overall brush height as a function of temperature, even for brushes with $h_{dry}$ as thin as 13 nm.

Importantly, the two-phase profiles obtained in the vicinity of the LCST, consistent with those extracted from neutron reflectometry\cite{Neutron3} or spectral ellipsometry\cite{collapseEllipso}, represent one of the very few direct experimental evidence for the vertical structuring of PNIPAM brushes upon collapse. Its physical origin can be traced back to the particular thermodynamic properties of PNIPAM/water mixtures. From their experimental study of the phase diagram of PNIPAM in solution in water, Afroze {\it et al.}\cite{Afroze} have shown that the measured coexistence curve could be reproduced semi-quantitatively by modifying the usual Flory-Huggins free energy of mixing in order to account for a concentration dependence of the interaction parameter, $\chi_{eff}(\phi)$, which characterizes the difference of interaction energies in the mixture. Such a $\phi$-dependence can be attributed to monomer/solvent hydrogen bonds\cite{2state}. It suggests that PNIPAM/water systems exhibit a so-called type II phase transition involving the coexistence of two phases having different but finite polymer concentrations\cite{Afroze,Avi4}. As a consequence, theoretical works on monodisperse polymer brushes have shown that $\chi_{eff}(\phi)$ could lead to density profiles exhibiting a vertical phase separation within the brush\cite{Ncluster,Avi1}. 

However, the effect of polydispersity on the sharpness of vertical phase separation has not, to the best of our knowledge, been studied theoretically. One can therefore wonder whether a finite polydispersity could be responsible for the vertical phase separation we observe experimentally. It is known from the behavior of free PNIPAM chains in aqueous solutions that longer chains tend to display a lower LCST than shorter ones, and hence collapse earlier in temperature\cite{length}. In a polydisperse brush, we thus anticipate that, at a given temperature close to the ``average'' LCST, shorter chains will remain solvated while longer ones will collapse. On this basis, vertical phase separation is expected to be smoothed out rather than enhanced by polydispersity, which is thus unlikely to be at the origin of the ``phase-separated'' profiles that we observe.

\begin{figure}[h!]
$$
\includegraphics[width=8cm]{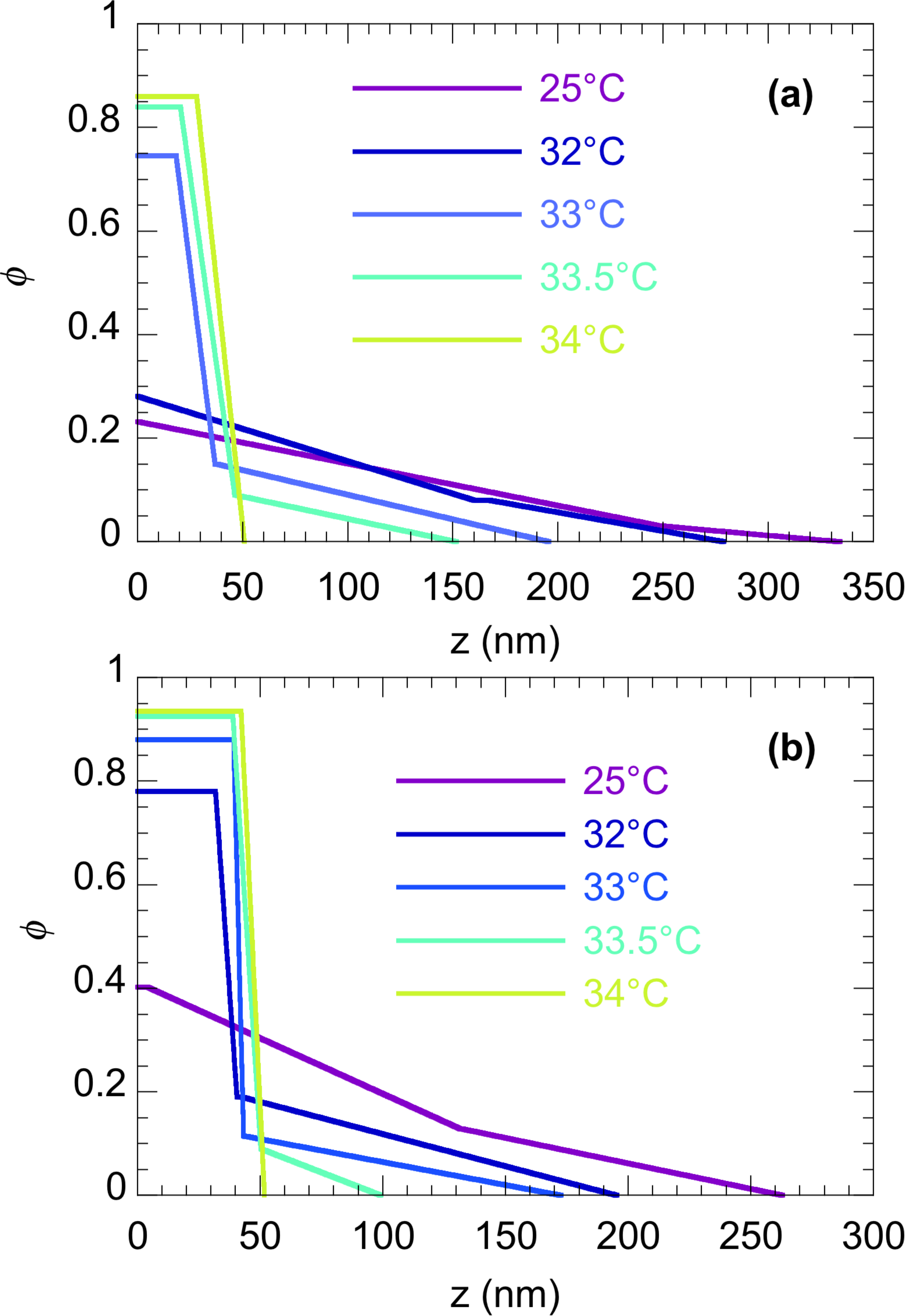}
$$
\caption{\label{fig:profilesLD-MD} Density profiles obtained from reflectance spectra (not shown) as a function of temperature for: (a), the low-density brush LD; and (b), the medium density brush MD.}
\end{figure}

We are therefore prompted, as proposed in previous studies\cite{Neutron3}, to associate the two-phase density profiles observed here
with the vertical phase separation predicted upon varying the solvent quality. In further support of this, $\phi(z)$ profiles obtained for the LD and MD brushes also point to vertical phase separation near the LCST (see Fig. \ref{fig:profilesLD-MD}), and display shapes that are qualitatively in good agreement with those computed from self-consistent field (SCF) theory using the phenomenological free energy of mixing of Afroze\cite{Avi2}. Besides, we remark that while vertical phase separation is clearly seen between 30 and 34$^{\circ}$C for the HD brushes, this range is reduced to 32-34$^{\circ}$C for the MD brush and to 33-34$^{\circ}$C for the LD one (Fig. \ref{fig:profilesLD-MD}). This indicates that vertical phase separation is affected by the grafting density, and suggests that it may disappear for low enough grafting densities, which is fully consistent with theoretical predictions\cite{Avi2,Avi1}.

A closer comparison however reveals that, while SCF theory predicts that vertical phase separation should be observed, for brushes of high grafting densities, at temperatures as low as 27$^{\circ}$C\cite{Avi2}, we observe experimentally that two-phase profiles are unambiguously found at $T\gtrsim 30^{\circ}$C only, as concluded from neutron reflectometry studies\cite{Neutron3}. This quantitative discrepancy is likely to result from the fact that SCF theory uses the empirical form of $\chi_{eff}$ determined by Afroze\cite{Afroze,Avi2}, which accounts only semi-quantitatively for the PNIPAM/water phase diagram. Such quantitative differences between theory and experiments call for further studies of this phase diagram in order to determine more accurately the $\phi$-dependence of $\chi_{eff}$, or, alternatively, for the use of data obtained on brushes, such as those presented here, to deduce $\chi_{eff}(\phi)$ from confrontation with theory.

Overall, the results presented in this section provide strong support for vertical phase separation upon collapse of PNIPAM brushes, with qualitative trends in excellent agreement with theoretical predictions.

\subsection{Brush swelling at room temperature}
\label{subsec:lowTemp}

\begin{figure}
$$
\includegraphics[width=8cm]{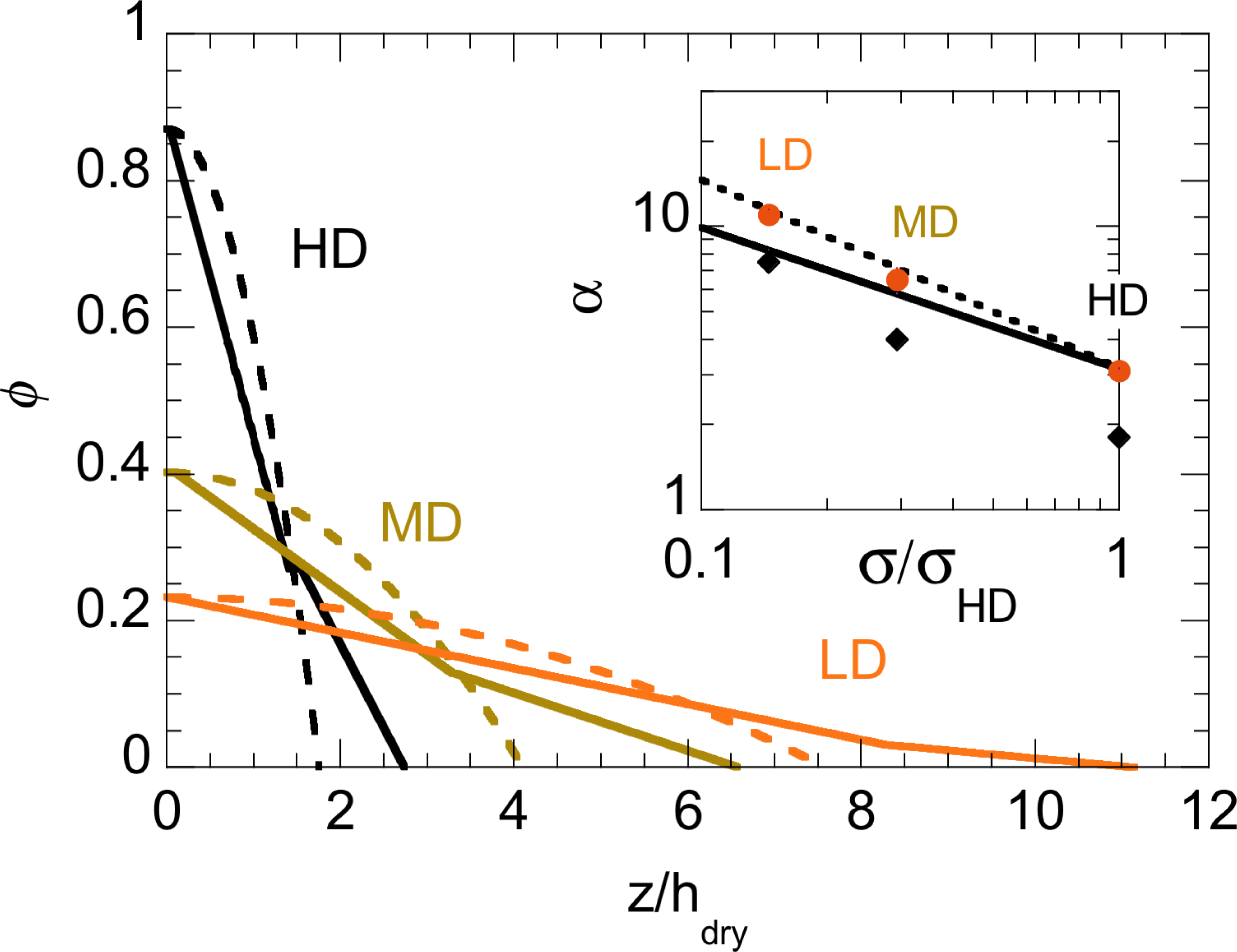}
$$
\caption{\label{fig:profiles25deg} Density profiles of PNIPAM brushes at 25$^{\circ}$C. Profiles retrieved from spectral reflectance measurements are plotted as solid lines for brushes HD4, MD and LD (see Table~\ref{table1}), along with calculated parabolic profiles corresponding to the same total amount of monomer and the same density at the grafting surface (dotted lines). Inset: Swelling ratio $\alpha$ as a function of grafting density (normalized by $\sigma_{HD}$). $\alpha$ determined from the density profiles extracted from data fits ($\bullet$), and from the parabolic profiles ({\small$\blacklozenge$}). The dependence of $\alpha$ on $\sigma$ expected from Eq. \ref{eq:sig} is plotted for comparison for $\nu=1/2$ (solid line) and $\nu=3/5$ (dotted line).}
\end{figure}

We discuss here the density profiles of the brushes obtained below the LCST, at $T=25^{\circ}$C. Such profiles are plotted in Fig. \ref{fig:profiles25deg} for the brush of high grafting density HD4 (see Table) and for the MD and LD brushes. For the three brushes, the monomer volume fraction $\phi(z)$ exhibits a piecewise linear decrease from a value $\phi_0$ at the substrate (with $\phi_0$ being lower for smaller grafting density) to 0 at a distance that we call $h_{swell}$ in the following, which corresponds to the height of the brush under good solvent condition.

If we compute the swelling ratio, $\alpha=h_{swell}/h_{dry}$, for the various brushes, we observe that: 

(i) $\alpha$ is consistently on the order of 3 for the HD brushes, indicating that brush swelling is fairly insensitive to chain length (see Table \ref{table1}),

(ii) $\alpha$ is larger for lower grafting densities, and increases from 3 for HD brushes to 6.5 for the MD sample and 11 for the LD one. Such values of $\alpha$ for the HD and MD brushes are in quantitative agreement with those obtained from a previous Surface Forces Apparatus study performed on samples of comparable grafting density \cite{bureau1}, as well as with recent numerical computations \cite{leonforte}. 

Theoretical predictions for $h_{swell}$ \cite{degennes,zhulina} allow us to obtain the following expression for the swelling ratio of a brush, as detailed in previous work\cite{bureau1}:
\begin{equation}
\label{eq:sig}
\alpha=\left(\frac{1}{a\sqrt{\sigma}}\right)^{3-1/\nu}
\end{equation}
with $\sigma$ the grafting density and $a$ the monomer size. The theoretical value of $\nu$ is 3/5 under good solvent conditions, but has been shown to fall closer to 1/2 from experiments performed with brushes of high grafting densities\cite{bureau1,nu}. From Eq. \ref{eq:sig}, one expects $\alpha$ to be independent of $N$, and to be larger for lower $\sigma$, which is fully consistent with our experimental observations. More precisely, Eq. \ref{eq:sig} predicts that a $P$-fold reduction of $\sigma$ induces a $P^{(3-1/\nu)/2}$-fold increase of the swelling ratio. Starting from the average value of the swelling ratio for the HD brushes, $\alpha\simeq 3.1$, and using the two limit values of $\nu$, we thus expect the swelling ratio of the MD brush, which is 3.5 times less dense, to be in the range $5.8-7.1$, while $\alpha$ for the LD brush, whose density is 7 times lower than the HD ones, should lie in the range $8.2-11.4$. The measured values for the MD ($\alpha$=6.5) and LD ($\alpha$=11) brushes are therefore in very good quantitative agreement with the above scaling analysis, as illustrated in Fig. \ref{fig:profiles25deg} (see inset). 

Such an agreement prompts us to estimate the grafting densities and chain lengths of our samples as follows: we use Eq. \ref{eq:sig} above, with $\nu=1/2$ and $a\simeq0.6$ nm\footnote{We estimate $a$ from $\rho=M_0/(N_Aa^3)$, with $\rho=0.95$ g.cm$^{-3}$ the NIPAM monomer density, $M_0=113$ g.mol$^{-1}$, and $N_A$ the Avogadro constant.}, to compute $\sigma_{HD}$ from $\alpha$ for the HD brushes. Using this value of the grafting density along with the measured dry thicknesses of the HD brushes, the number of monomer per chain, $N$, can be estimated from Eq. \ref{eq:1} (see Materials and Methods). The grafting densities of the MD and LD brushes are computed by dividing $\sigma_{HD}$ respectively by 3.5 and 7, as suggested by ellipsometry measurements, and chain lengths are subsequently estimated from Eq. \ref{eq:1}. This procedure provides estimates of the characteristics of the various brushes studied, as reported in Table \ref{table1}. It allows us to compute the radius of the coil that the chains would form if they were free in solution and non-interacting, and to build the ratio of such a radius to the average distance between grafting points. As shown in Table \ref{table1}, this ratio is found to be larger than 1 for all of our samples, both in the swollen and collapsed state. This points to the fact that the studied samples are well in the brush regime at all temperatures, including far above the LCST where water is a poor solvent. Under such conditions of strong overlapping, it has been shown by numerical simulations\cite{grest} that possible in-plane heterogeneities, the so-called octopus micelles due to lateral chain aggregation, are not expected to build up upon collapse. The studied brushes are therefore expected to remain laterally homogeneous as temperature is varied across the LCST, which ensures that the results presented in this work are not biased by any smoothening over small-scale lateral heterogeneities. The density profiles reported here thus reflect the actual axial chain conformation within the brushes, statistically averaged over the number of chains present within the probed area (the lateral resolution of our setup in its present configuration is on the order of 500 nm).

The above discussion is complemented by analyzing further the shapes of the density profiles presented in Fig. \ref{fig:profiles25deg}. The brush profiles clearly deviate from the expected parabola predicted both by theory\cite{milner} and simulation\cite{binder} under good solvent conditions. However, such predictions correspond to brushes made of monodisperse chains, and it has been shown by self-consistent field theory \cite{milner,devos} that this parabolic shape is lost when chain polydispersity is accounted for: when compared to the parabolic decay expected for monodisperse chains, the profile of a polydisperse brush, starting at the same monomer density at the wall, is predicted  to adopt a concave shape and to drop at $\phi=0$ further away from the grafted surface\cite{milner,devos}. Therefore, we have computed, for the three grafting densities investigated in this study, the parabolic profiles $\phi(z)=\phi_0(1-(z/h_{swell})^2)$ satisfying the mass conservation constraint, namely $\int_{0}^{h_{swell}}\phi(z)dz=h_{dry}$, and starting at the same $\phi_0$ as that determined from fitting the experimental data. Under such conditions, parabolic profiles are entirely determined, without any free parameter. A comparison between such parabolic profiles and those determined by fitting the experimental data using our model is provided on Fig. \ref{fig:profiles25deg}. In good agreement with the above-described theoretical predictions, we see that the $\phi(z)$ profiles we extract from our reflectivity data display the concave shape and tailing that characterize polydisperse brushes. In particular, we observe that the profiles extracted from the data fall at $\phi=0$ at a distance $z$ larger than that of the parabolic decays, {\it i.e.} the brushes swell more than expected for monodisperse chains, by about 50-60\% for the three grafting densities ($\alpha^{poly}/\alpha^{mono}\simeq 1.5-1.6$, see inset of Fig. \ref{fig:profiles25deg}). The magnitude of such a polydispersity-induced overswelling is theoretically predicted to depend on the polydispersity index\cite{devos}, defined as the ratio of the weight- to the number-averaged molar mass of the chains, $M_w/M_n$. Therefore, we conclude that our brushes are polydisperse, but possess consistent polydispersity indices from batch to batch, as indicated by the fact that brushes of various $\sigma$ display the same amount of overswelling. If we use the theoretical results of de Vos and Leermakers \cite{devos}, who have computed $\alpha^{poly}/\alpha^{mono}$ as a function of $M_w/M_n$, we find that an overswelling of 50-60\% is predicted to correspond to $M_w/M_n\simeq 1.25-1.3$, which provides us with a value of the polydispersity index for our samples that is consistent with that reported for long PNIPAM chains grown using the same radical polymerization technique\cite{ARGET1}.

It is important to remark that, since measured swelling ratios are larger than for monodisperse brushes of the same grafting density and chain length, our use of Eq.  \ref{eq:sig} above leads us to underestimate $\sigma$ by a factor of about 2--2.5, and in turn to overestimate $N$  by the same factor.

To close this section, we note that, while brushes elaborated by the grafting-from technique are increasingly used in a variety of applications, their characterization in terms of grafting density and molecular weight is still a challenging task\cite{revBrushAppli2}. The most reliable technique consists of degrafting the tethered chains from the surface and subsequently analyzing their size and distribution by chromatography\cite{degraft}. This, however, requires the collecting of a large enough amount of polymer in order to perform an accurate solution characterization, and is therefore better suited for brushes grafted on samples having large surfaces rather than on flat substrates. Alternatively, polymerization is commonly performed, in parallel with brush growth, using free initiator in solution\cite{nu}. This strategy allows characterization of the free chains, and assumes that polymerization in the bulk and from the surface obey the same kinetics. However, such an assumption has been shown to break down\cite{degraft2} due to surface-induced crowding\cite{genzer2} and termination effects\cite{genzer1}. Another technique consists of evaluating the grafting density using a surface sensitive technique, such as X-ray photoelectron spectroscopy in order to determine the surface coverage of polymerization initiator, but relies on an assumption regarding the fraction of initiator that actually leads to chain growth\cite{LeckbandProt2}. The various means for determining the molecular parameters of grafted-from brushes thus present limitations in terms of ease of use or quantitative reliability. In this context, the analysis we propose here comes as an interesting alternative for brush characterization, as it permits evaluating, directly from the data of interest, \textit{in situ}, non destructively and without the need for additional bulk solution measurements, not only $\sigma$ and $N$, but also $M_w/M_n$ and the errors associated with the assumptions on which the analysis relies.

\subsection{Hysteresis}
\label{subsec:hysteresis}

\begin{figure}
$$
\includegraphics[width=8cm]{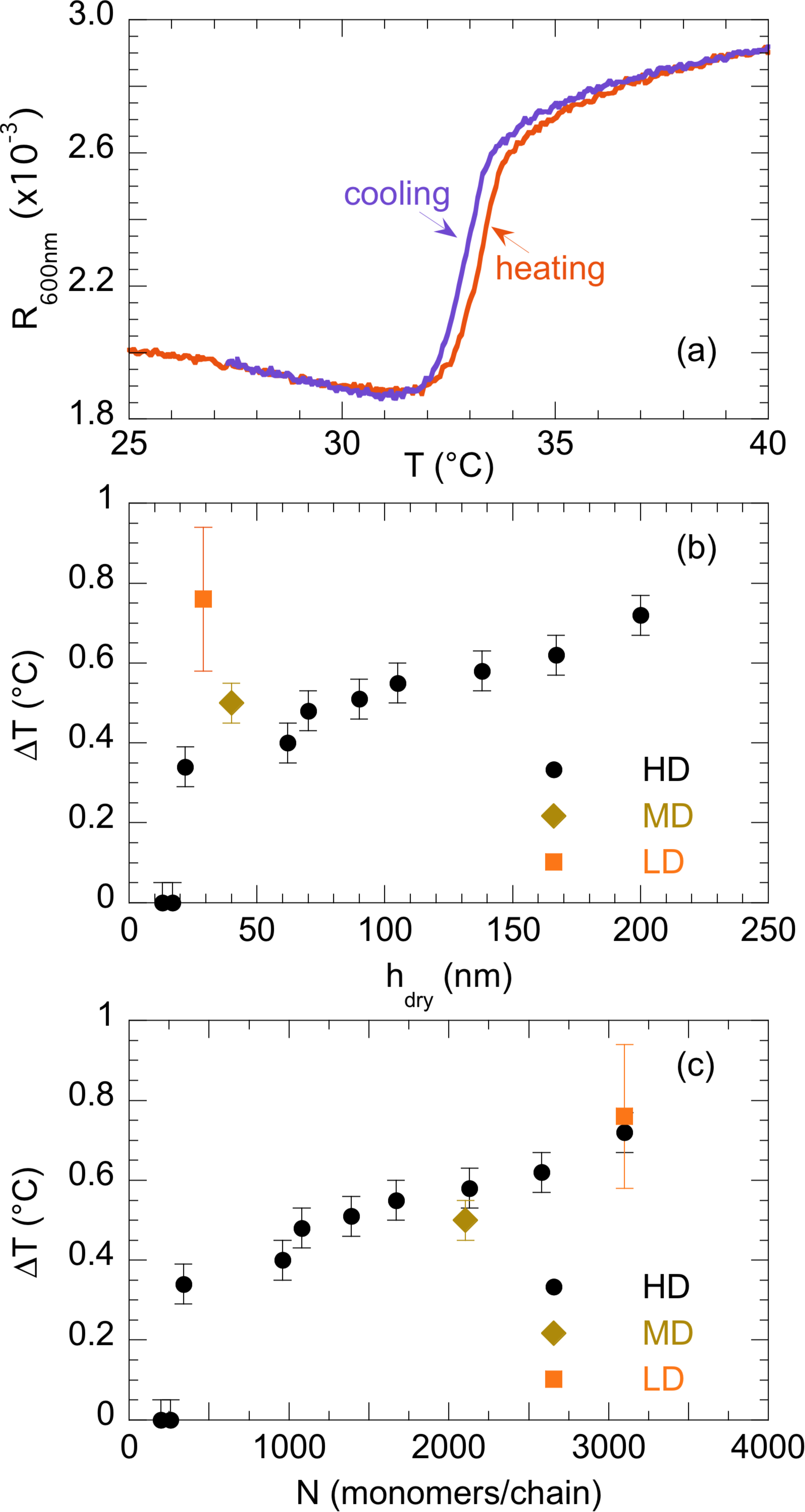}
$$
\caption{\label{fig:hysteresis} Hysteresis in the thermal conformational change of PNIPAM brushes. (a), example of reflection profile at $\lambda=600$nm upon heating (red) and subsequent cooling (blue) for the medium density brush MD, illustrating the shift in transition temperature depending on the thermal history of the brush. Hysteresis amplitude $\Delta T$ is plotted for the brushes listed in Table~\ref{table1} as function of (b), their dry thickness $h_{dry}$; and (c), the chain length. Error bars in (b) and (c) are estimated for the dispersion of the measured values of $\Delta T$ at different temperatures and different wavelengths.}
\end{figure}

For most of the brushes studied here, there is a range of temperature around the LCST, typically between 31 and 34$^{\circ}$C, in which we observe a clear difference in the reflectivity spectra measured at the same temperature upon heating and cooling. This hysteretic response is illustrated in Fig. \ref{fig:hysteresis}(a), showing, for the MD brush, the reflectivity measured at a wavelength of 600 nm during a heating/cooling cycle. We have quantified the hysteresis in two different ways: (i) from reflectivity $vs$ temperature curves, such as that shown in Fig. \ref{fig:hysteresis}(a), we have computed the shift in temperature, $\Delta T$, required in order to overlap the cooling curve with the heating curve in the transition region, (ii) we have measured the difference between the two temperatures at which the reflectivity spectra upon heating and cooling are identical over the full spectral range (this was done at three different reference temperatures in the range 31-34$^{\circ}$C). These two methods yielded consistent values for the temperature hysteresis $\Delta T$. 

In Fig. \ref{fig:hysteresis}b $\Delta T$ is plotted as a function of the brush dry thickness $h_{dry}$, for all the brushes elaborated. The series of HD brushes display the following trend: below a threshold thickness of $h_{dry}\simeq 15$ nm, no hysteresis is observed, whereas thicker brushes exhibit a hysteresis that increases from 0.3 up to 0.75$^{\circ}$C as $h_{dry}$ is increased. However, we observe that, on such a $\Delta T(h_{dry})$ plot, the results obtained for the MD and LD brushes lie well above those obtained for the high density samples. From the latter observation, we can conclude that the magnitude of the hysteresis is not simply controlled by $h_{dry}$, {\it i.e.} is not merely a function of the amount of monomer bound to the surface.

Now, if we plot the same data as a function of the average number of monomers per chain ($N$, as determined previously from the swelling ratio) for each brush, as shown in Fig. \ref{fig:hysteresis}(c), we observe that the results for the LD and MD samples are horizontally shifted and closely follow the trend obeyed by those obtained for the HD brushes: below a threshold chain length, no hysteresis is observed, while above this threshold $\Delta T$ increases with $N$, independently of the brush grafting density. Such a collapse of the data thus indicates that the hysteresis observed in one heating/cooling cycle is essentially governed by the length of the grafted chains.

Hysteresis in the thermal response of PNIPAM brushes has already been reported in several previous works using SPR and QCM  \cite{QCM1,QCM2,QCM3,QCM4,SPR}. Two possible mechanisms have been proposed to explain this phenomenon \cite{QCM1,QCM2,QCM3,QCM4}: (i) the formation, in the collapsed state above the LCST, of inter-monomer hydrogen bonds that act as ``crosslinks'' and must be disrupted upon reswelling of the chains, as suggested by experiments performed on PNIPAM chains in solution\cite{Wu1}, and (ii) the presence of entanglements that create transient  topological constraints that limit the extent of chain swelling, thus delaying the swelling kinetics as the temperature is decreased across the LCST. The first mechanism attributes hysteresis to the difference in energy barrier between the collapse and swelling processes, while the second one is rather a kinetic scenario in which the hysteresis is mainly governed by the time scale required to relax the topological constraints due to entanglements compared to the time scale at which the temperature is decreased. Both mechanisms have been proposed on a qualitative basis, and no systematic studies of hysteresis as a function of brush parameters have yet allowed discrimination between the two.

The results presented here provide new insights into the origin of the observed hysteresis. Indeed, if inter-monomer H-bonds were entirely responsible for hysteresis, one would expect $\Delta T$ to be directly proportional to the amount of monomers, hence to $h_{dry}$, with $\Delta T$ smoothly vanishing to zero as $h_{dry}$ tends to zero, in contrast to the threshold observed in our data (Fig.\ref{fig:hysteresis}~(b)). Moreover, such a mechanism cannot account for the fact that, at a given $h_{dry}$, {\it i.e.} for the same amount of monomers, hysteresis is larger for brushes of lower grafting densities. 

On the other hand, the second mechanism, which proposes entanglements as the main source of hysteresis is fully consistent with our results: entanglements can form only when polymer chains are longer than a critical length (or, equivalently, have a molecular mass larger than the critical mass for entanglement, $M_c$).  Such a critical length accounts for the observed threshold in the $\Delta T(N)$ curve shown in Fig. \ref{fig:hysteresis}(c). More precisely, we find the threshold to be around $N_c$=300 monomers/chain. Taking into consideration the above discussion regarding our estimates for $N$, that can be too large by a factor of 2, let us  consider conservatively that $N_c$ is in the range 150--300. This translates into a critical mass $M_c=N_cM_0$ in the range 17--34 kg.mol$^{-1}$, which covers the reported value of PNIPAM entanglement mass of 23 kg.mol$^{-1}$ \cite{Mc}.

Moreover, increasing the density of entanglements in a semi-dilute solution of linear chains is known to slow down their relaxation \cite{Wu3}. In a similar way, one therefore expects that a higher density of entanglements within a brush will delay further its reswelling, thus yielding a larger apparent hysteresis at a given rate of temperature change. This is in agreement with our results showing that $\Delta T$ is larger for longer macromolecules, for which the number of entanglements per chain increases.

We further validate this picture by comparing the lifetime of entanglements with the time scale of temperature change in our experiments. The order of magnitude of the latter, $\tau_{exp}$, is given by the time it takes to change the temperature by about 1$^{\circ}$C around the transition. With the rate of temperature decrease being fixed to approximately 0.15$^{\circ}$C.min$^{-1}$ in all our experiments, this yields $\tau_{exp}\sim 400$ s. The lifetime of entanglements, on the other hand, depends on the relaxation mechanism by which constraints are released. In contrast with free linear chains in a solution or a melt, which can reptate in order to relax topological constraints, chains inside a polymer brush are bound by one end to a substrate. This implies that the only process by which a tethered chain can relax entanglements and explore its environment over lengthscales on the order of its size is the so-called arm-retraction mechanism. Such a mechanism, known to govern the dynamics of star or branched polymers \cite{HP}, describes the fact that a chain having one end fixed in space can escape the ``tube'' formed by entanglements and neighboring chains solely by retracting part or all of its length along this tube. This is associated to a large entropic penalty and a low probability of occurrence, such that the relaxation time of a tethered chain grows exponentially with the number of entanglements per chain \cite{HP,Vega,Ajdari}:
\begin{equation}
\label{eq:arm}
\tau_{arm}(s)\simeq \tau_0 N^2\exp\left(\mu s^2 N/N_e\right)
\end{equation}
where $\tau_{arm}(s)$ is the time to retract a chain made of $N$ units by a fraction $s$ ($0<s<1$) along the tube length, $N_e\sim M_c/(2M_0)\sim 100$  is the number of monomers between entanglements\footnote{We consider here that $N_e$ is approximately equal to half the critical length for entanglements $N_c$, as suggested from rheology studies on polymer melts\cite{Colby}, and use $M_c=23$ kg.mol$^{-1}$}, the constant $\mu\sim 15/8$, and the pre-exponential time $\tau_0 N^2$ is the Rouse time of the equivalent ``free'' chain \cite{Vega,Ajdari}. From the work of Yuan et al. \cite{Wu3}, we can estimate the Rouse time of PNIPAM chains of length $N\simeq 1800$ units in solution in water at 25$^{\circ}$C to be on the order of $6\times 10^{-3}$ s.

Thus, it can be estimated that, for $N=1800$, $\tau_{arm}(s)$ is comparable to or larger than $\tau_{exp}\sim 400$ s for any $s\gtrsim 0.55$, and reaches up to 10$^{12}$ s for full chain retraction ($s=1$). This is to be taken as an order of magnitude estimate only, as the value of $\mu$ or the exact pre-exponential factor are not firmly established\cite{Ajdari,Colby} (for instance, we have used a Rouse time determined at room temperature, whereas this pre-exponential time might actually be much larger since the chain dynamics above the LCST, in poor solvent, is expected to be significantly slower). However, this allows us to draw the following conclusions:

(i) entanglements relaxed by arm-retraction can indeed have a lifetime that is comparable to the experimental time scale,

(ii) the observed hysteresis thus most likely results from the contribution of those topological constraints that are relaxed by sub-chain rearrangements at the proper lengthscale ({\it i.e.} involving $s$ such that $\tau_{arm}(s)\sim \tau_{exp}$).

Overall, our results and their above analysis strongly support the fact that the hysteretic response of the PNIPAM brushes reported here has essentially a kinetic origin, and results from long-lived entanglements that delay swelling, as proposed theoretically in the description of the swelling dynamics of collapsed globules\cite{Johner}.
\begin{figure}[h!]
$$
\includegraphics[width=8cm]{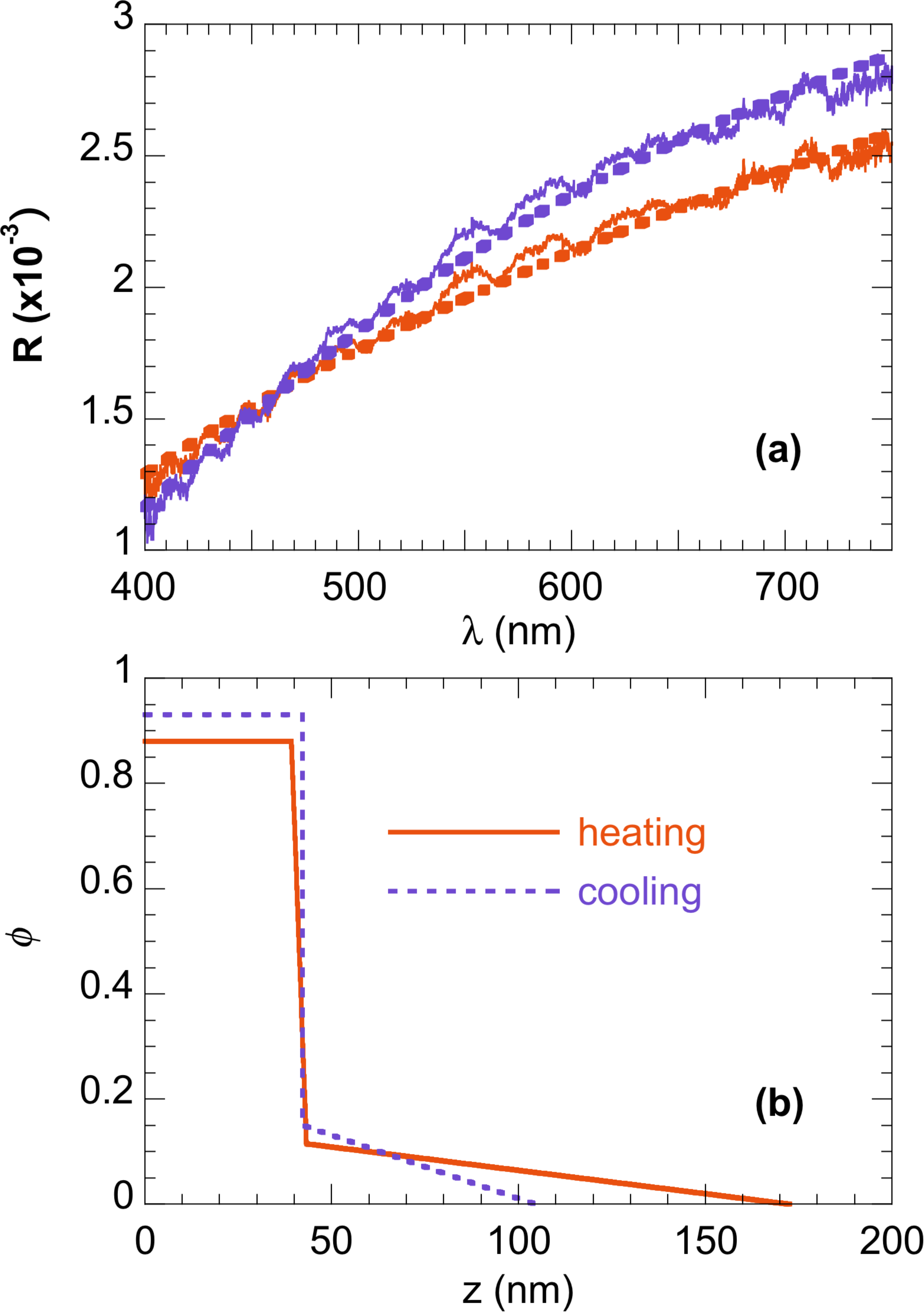}
$$
\caption{\label{fig:hysteresis2} (a) spectral reflectance and (b) extracted density profile of the medium density brush at 33$^{\circ}$C upon heating and cooling.}
\end{figure}

We conclude this section by coming back to the vertical structure exhibited by the brushes in the transition region. We have shown in the previous section that, upon heating across the LCST, brushes display a non-uniform density profile and exhibit a dense phase close the grafted substrate, surmounted by a more dilute layer. Analysis of the reflectivity spectra measured during cooling leads to similar non-uniform density profiles, as illustrated in Fig. \ref{fig:hysteresis2} for the MD brush. However, when comparing the profiles extracted from a fit of the spectra measured at the same temperature, in the hysteresis region, we observe that a brush displays, when cooled, a denser inner region and a less extended outer layer than when heated. This is in good agreement with the above picture describing hindered swelling of the brushes upon cooling. Moreover, this provides, to the best of our knowledge, the first direct experimental evidence that a brush can exhibit two different non-uniform structures when crossing the LCST from the swollen and from the collapsed state, in agreement with qualitative pictures proposed in previous works on brushes\cite{QCM4} or globules\cite{Wu1}.

\section{Conclusions}

We have shown that spectral reflectance measurements performed over the visible spectrum provide detailed information regarding the thickness, density and vertical structure of thin organic layers such as polymer brushes immersed in a solvent. Therefore, this technique offers an interesting alternative to Neutron Reflectivity or Spectral Ellipsometry, since it allows fast and in-depth characterization of brushes. Moreover, it presents several other advantages, which make it versatile and powerful:

(i) the setup is based on a microscope, enabling easy combination with other imaging techniques. It also allows for spatially-resolved reflectance measurements, which opens the way to mapping of thin film heterogeneities. 

(ii) optical reflectance measurements are performed on glass substrates, whereas NR or SE typically require silicon wafers. In the context of smart substrates for biology applications, which need to be transparent in order to be compatible with usual microscopy techniques, this permits working directly with the samples of interest for the target application, for {\it a priori} and {\it in situ} characterization before use.

We have applied this technique to study the effect of molecular parameters on the thermal response of brushes made of PNIPAM, which are  ubiquitous smart coatings used in sensing, switching or actuating applications. Our results provide an unprecedented support to recent theoretical predictions regarding the occurrence of vertical phase separation in PNIPAM brushes in water, and shed light on the physical mechanism underlying the hysteretic response of such brushes upon cycling in temperature.

\section{Aknowledgments}

We thank Ralf Richter and Claude Verdier for fruitful discussions, and Heather Davies for critical reading of the manuscript.
We acknowledge financial support from the Nanosciences Foundation in Grenoble, the Universit\'e Grenoble 1 ``AGIR'' program, the CNRS ``D\'efi Emergence'' program, and from the French Agence Nationale de la Recherche (ANR), under grant ANR-13-JS08-0002-01 (project SPOC). D.D. is part of the LabEx Tec 21 (Investissements d'Avenir - grant agreement n$^{\circ}$ANR-11-LABX-0030).


\section{Materials and Methods}

\subsection{Synthesis and characterization of brushes}
\label{subsec:chemistry}

The series of 12 brushes, whose characteristics are provided in Table \ref{table1}, was prepared by surface-initiated Atom Transfer Radical Polymerization, as described in SI.

Brushes were grafted in parallel (same reaction batch) on a glass coverslip, used for reflectivity measurements, and on a piece of Si wafer, which we used for the determination of the dry brush thickness by ellipsometry. Ellipsometry measurements were performed on a rotating quarterwave plate home-built instrument working at 70$^{\circ}$ incidence angle and 632 nm wavelength. Data analysis was done assuming a multilayer model including a silicon substrate, a silicon oxide layer (thickness of 2 nm and refractive index 1.46), and an outermost polymer layer of refractive index 1.46 and thickness to be determined. Each sample was measured at 3-5 different spots on its surface, yielding consistent values of the brush thickness to within $\pm$1 nm.

For the low and medium density (LD and MD) samples, PNIPAM polymerization was performed in parallel on a coverslip and a wafer with the same initiator density, as well as on a control wafer of maximum density (HD). The dry thickness of a brush ($h_{dry}$), as determined by ellipsometry, is related to the grafting density ($\sigma$), the number of monomer per chain ($N$) and the monomer volume ($a^3$) by:
\begin{equation}
\frac{h_{dry}}{\sigma}=Na^3
\label{eq:1}
\end{equation}
Samples of different $\sigma$ grown under the same polymerization conditions should exhibit a similar $N$, and differences in $h_{dry}$ observed between the lower density substrate and the maximum density control mostly reflect the variations of $\sigma$. Namely, the MD brush was measured to have a dry thickness of $h_{dry}^{med}=40$ nm, while the full density sample had a thickness of $h_{dry}^{HD}=140$ nm, and the LD brush displayed $h_{dry}^{low}=29$ nm, for $h_{dry}^{HD}=200$ nm. We thus conclude that the MD brush is 3.5 times less dense than the HD, while the LD brush exhibits an almost 7-fold decrease in grafting density.

\subsubsection{Pre-measurement protocol}
\label{subsubsec:protocol}

Before use, the brush-grafted substrates were left to soak in ultrapure water at room temperature overnight  to remove any weakly physisorbed polymer from the surfaces. The samples were then dried, and the brush dry thickness re-measured by ellipsometry (typically, brushes were found to exhibit a small decrease in thickness, between 1 nm for the thinnest ones and 3-5 nm for the thickest. Values reported in Table \ref{table1} correspond to this post-soaking measurements). 

The brush-bearing glass coverslips were then mechanically scratched with a blunt steel blade to remove the brush from the surface over a narrow region (a few tens of micrometers in width) along the coverslip diameter. This provided a glass/water interface devoid of brush that we used as a reflectivity reference, as explained in the next section. 

Coverslips were then carefully rinsed with water, dried, and mounted on a custom-built holder designed as a liquid cell. A mica sheet of 20-25$\mu$m thickness was glued to the bottom of the coverslip using UV-curing optical glue (see next section). The cell was filled with ultrapure water, sealed, and put at 4$^{\circ}$C for two hours in order to ``reset'' the thermal history of the brushes. Afterwards, the holder was installed on the microscope stage, and allowed to come back to room temperature before data acquisition.

\subsection{Reflectance measurements}
\label{subsec:RICM}

\subsubsection{Experimental setup}
\label{subsubsec:setup}


\begin{figure}
$$
\includegraphics[width=8cm]{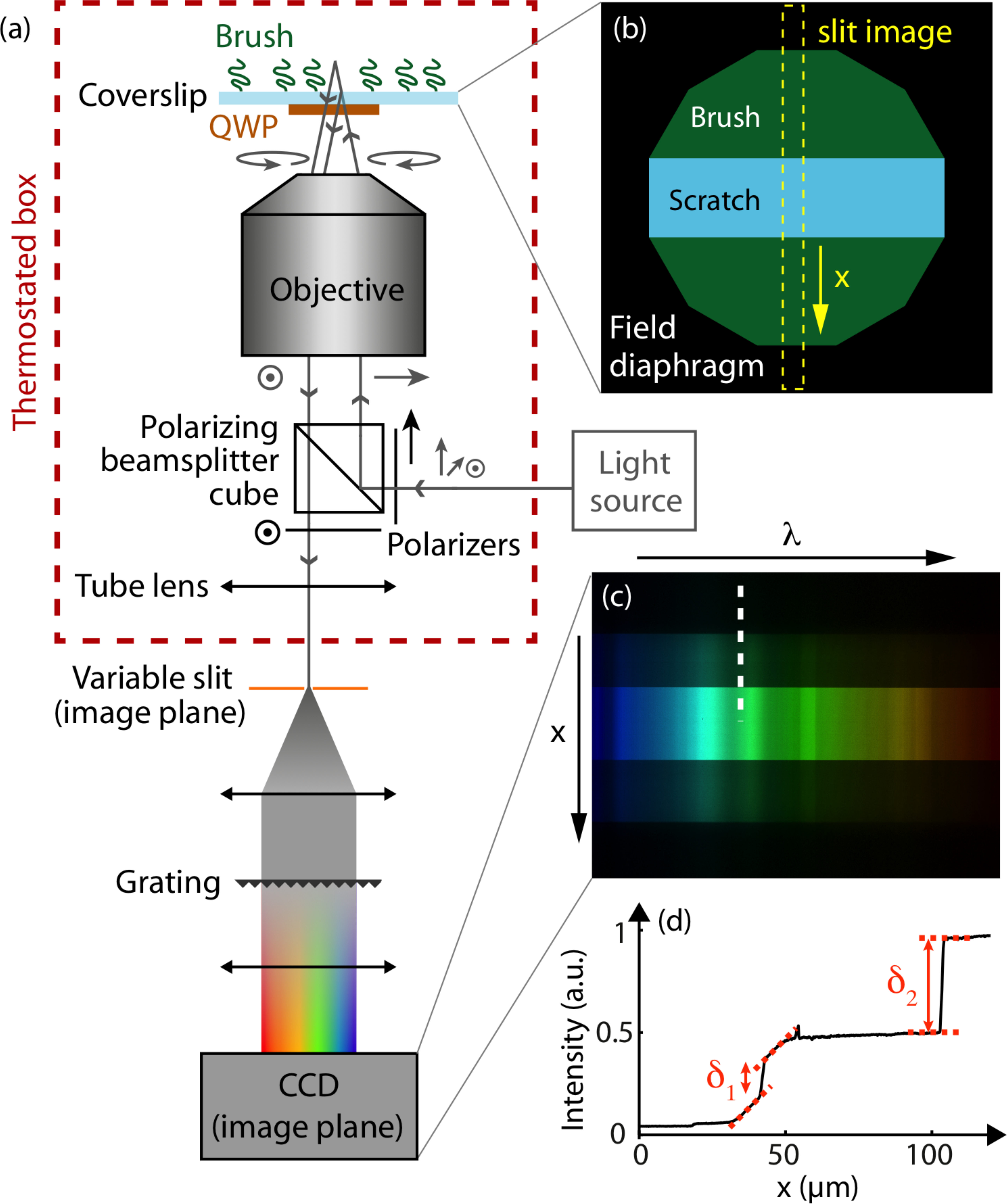}
$$
\caption{\label{fig:setup}(a), schematic view of the RICM microscope used for spectral reflectance measurements. QWP, quarter waveplate. (b), schematic view of a sample image showing the position of the slit (in orange in (a)) on the image. (c), chromatically dispersed image of the slit in (b), recorded on the CCD. (d), profile along the dotted white line in (c) showing the quantities extracted on the images to calculate the spectral reflectance.}
\end{figure}

A scheme of our microscope is shown on Fig. \ref{fig:setup}. Briefly, a plasma light source (Thorlabs) emitting a continuous spectrum in the range 400-800nm was linearly polarized and  coupled to the microscope using a polarising beamsplitter cube (Thorlabs). The illumination numerical aperture (INA) was set to 0.15 using a calibrated aperture diaphragm. The reflected light was detected through the beamsplitter cube and a crossed polarizer. 

In typical RICM systems, imaging is achieved using an Antiflex objective (Zeiss, 63X, 1.3NA) that upon double pass induces a 90$^{\circ}$ rotation of the polarisation\cite{RICM}. Here we introduce an alternative method for collecting reflected signals through cross polarisers: by gluing a mica sheet of controlled thickness to the bottom of the sample, we achieve polarisation control  without using a specifically designed objective. In the following experiments, thicknesses of 20-25$\mu$m were used, corresponding to a quarter-wave-plate in the 450nm range. This approach allows the use of objectives of any magnification, immersion medium and numerical aperture. Here, a 20X, 0.75NA super apochromat air objective (UPLSAPO 20X, Olympus) was used to image the sample without thermal coupling through the immersion medium. The reduced chromatic aberrations ensure good imaging quality throughout the visible spectrum.

A home-built spectrometer was coupled to the output port of the microscope for spectral reflectance measurements: a variable slit placed at the imaging plane was used to control the spectral resolution and was imaged using achromatic lenses onto a CCD camera after reflection on a diffraction grating (all Thorlabs). A spectral image of a line in the sample was thus obtained on the camera (Fig. \ref{fig:setup}c).

Finally, an autofocus device coupled to the axial motorisation of the microscope (ASI) allowed long-term data acquisition during temperature changes without losing the focus. Because this device is based on the use of a 785nm diode, however, spectral acquisition of the data was limited to up to 750nm in wavelength.

Temperature control was achieved using a thermostated box (Digital Pixel) enclosing most of the microscope. Such an arrangement led to slow temperature variations during heating (0.3$^{\circ}$C/min) and cool down (0.15$^{\circ}$C/min) and ensured that thermal gradients in the sample were reduced to a minimum during the experiment, thereby allowing unbiaised measurements of the brush temperature. 

Furthermore, temperature measurement was performed as close as possible to the imaged region of the brush. This was done by using a miniature type-K thermocouple encased within the liquid cell holder, with the sensing region resting on the brush-grafted surface. A digital multimeter (Agilent 34970A) was used, in conjunction with a second sensor measuring the reference temperature of the thermocouple cold junction, in order to monitor the temperature of the sample with an absolute accuracy of $\sim$0.1$^{\circ}$C and a resolution of about 0.01$^{\circ}$C.

A custom-written software (LabView) was used to synchronize illumination of the sample and image acquisition with temperature measurements at fixed temperature intervals of 0.05$^{\circ}$C.

\subsubsection{Spectral reflectance measurements}
\label{subsubsec:spectref}

Spectral reflectance at near normal incidence is a well-established technique for {\it in situ} characterisation of thin metallic\cite{ Killeen:1994ws} or polymer\cite{ Gauglitz:1993tb} films. Here we expand this method by performing spectral reflectance measurements with a microscope: we thus obtain a diffraction-limited lateral resolution instead of an averaged measurement on a surface,  opening the way to sub-micron mapping of thin layer heterogeneities. 

In order to measure absolute spectral reflectances, the sample was placed on the microscope such that the scratch on the brush was located in the middle of the field of view, so that it appeared as a line in the spectral dimension on the CCD. In addition, the field diaphragm of the microscope was partially closed so that its edges were visible on the CCD as dark regions on either sides of the image (Fig. \ref{fig:setup}). In this configuration, we could simultaneously measure on the same image a ``dark" reference (using the field diaphragm image) and a ``bright" reference (using the bare region in the scratch).

Each measurement was typically performed using an exposure time of 0.5-1s, limited by the signal intensity due to the small value of the reflectance of thin polymer brushes.

\subsection{Data analysis}
\label{subsec:analysis}

\subsubsection{Image processing}
\label{subsubsec:imgproc}

\begin{figure}
$$
\includegraphics[width=8cm]{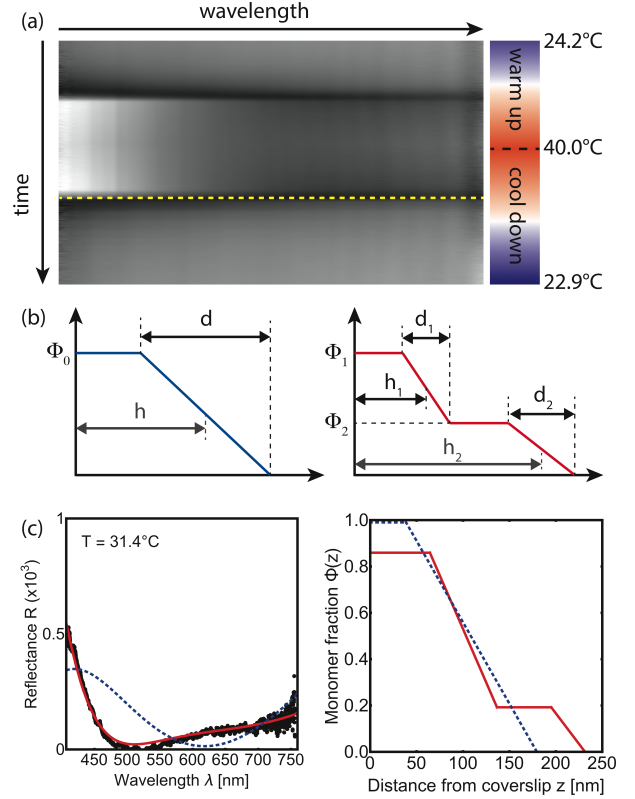}
$$
\caption{\label{fig:spectra2D}Extraction of the monomer density profiles from spectral reflectance curves. (a), example of spectral reflectance data, $R(\lambda, \, t)$, obtained after processing of a warm-up/cool-down cycle. (b), schematic representation of the one-phase (left) and two-phase (right) profile, showing the parameters used to describe each profile. Only parameters in black are independent and used to fit the profile, while parameters in gray are deduced from the conservation of the total amount of monomer (see text and Eqs. \ref{equreflgen}-\ref{equrefl3}). Parameters $h$, $h_1$ and $h_2$ correspond to the distance from the wall ($z=0$) to the midpoint of the linear decay(s). (c), example of fit around the LCST for sample HD4. Left panel: experimental data (black dots) corresponding to the intensity profile along the horizontal dotted line drawn in (a), along with the best fits obtained, according to the procedure described in the text, using a one-phase (dotted blue lines) or a two-phase (red solid line) profile. This illustrate that a two-phase profile is mandatory to properly fit the data around the LCST. Right panel: corresponding one- and two-phase profiles.}
\end{figure}

Image processing was achieved using home-written ImageJ macros. The reflectivity of the brush covered glass was calculated with respect to that of the bare glass in water. Using both a ``dark" and a ``bright" reference allows light source fluctuations and background noise caused by spurious reflections (top surface of the sample holder, mica/glass interface, etc.) to be accounted for. In practice, two values can be extracted from the RICM image for each wavelength: $\delta_1=I_{brush}-I_{black}$, and $\delta_2=I_{bright}-I_{brush}$ (Fig. \ref{fig:setup}d). These quantities were calculated for each wavelength and temperature point, resulting in 2D temperature-wavelength maps (Fig. \ref{fig:spectra2D}a). Noise was reduced by filtering these maps with a 3x3 median kernel, and reflectance of the sample relative to that of the bare glass/water interface was subsequently computed as $R(\lambda,t)/R_{glass/water} (\lambda,t)=(I_{brush}-I_{black})/(I_{bright}-I_{black})=\delta_1/(\delta_1+\delta_2)$. Since the refractive indices of glass and water are known, the absolute reflectance of the sample is straightforwardly obtained.

\subsubsection{Spectral reflectance curve fitting}
\label{subsubsec:curvefit}

\begin{figure}[h!]
\includegraphics[width=8cm]{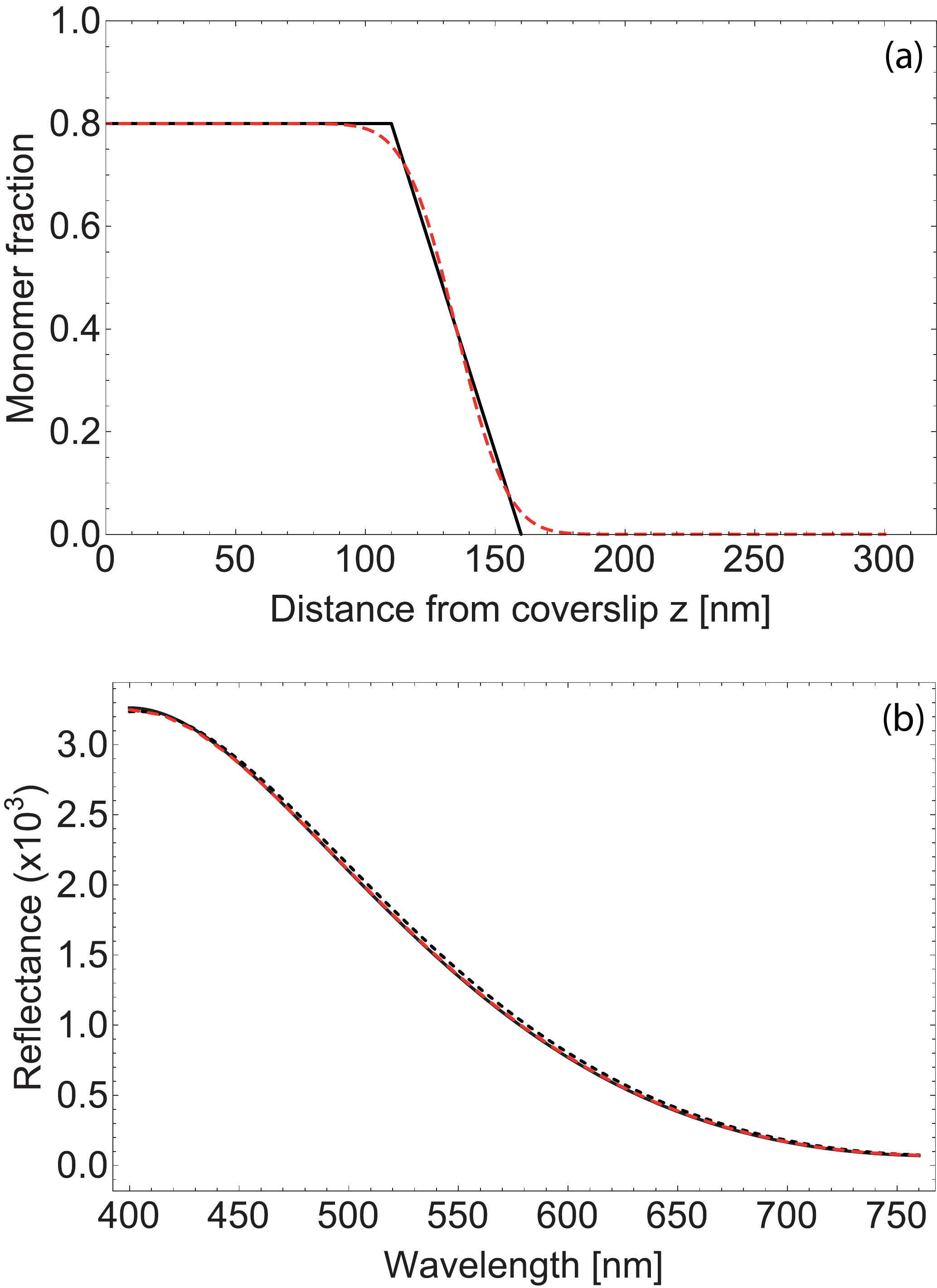}
\caption{\label{fig:comparisonapproximations} (a), two one-phase profiles with identical amounts of monomer, and differing by their smoothness, are used to derive the spectral reflectance curves shown in (b) (black solid line and red dashed line). The straight-line profile was also used to calculate the exact reflectance without using the Born approximation (black dotted line). The three curves do not show any significant difference.}
\end{figure}

The 2D temperature-wavelength reflectance maps were analyzed in two ways: first, the reflectance profile as a function of temperature can be extracted for a given wavelength, e.g. to analyze the hysteresis observed during a warm up/cool down cycle. Conversely, computation of the axial density profile of the brush is achieved by fitting individual spectral reflectance curves. To this aim, we introduce several simplifying hypothesis:

\begin{itemize}
\item Each reflecting layer in the sample has a small reflection coefficient, so that the total reflection can be approximated by the integral of reflections of infinitesimal homogeneous axial slices. The reflectance can then be written as:
\begin{widetext}
\begin{equation}
R(\lambda)=\left\lvert\frac{n_{glass}-n(0)}{n_{glass}+n(0)}+ \right.
\left. \int^{+\infty}_{0}\!\!\!\!\!\frac{1}{2 n(z)}\frac{\partial n}{\partial z}\exp(\frac{2\pi}{\lambda}\int^{z}_{0}n(z')dz')dz\right\rvert^2
\end{equation}
\end{widetext}
where $n(z)$ is the refractive index of the polymer brush at axial position $z$ within the brush, and $n_{glass}$ is the refractive index of the glass coverslip (BK7).
\item The refractive index of the hydrated polymer varies linearly with the monomer volume fraction $\phi (z): n(z)=n_{NIPAM}*\phi (z)+n_{water}*(1-\phi (z))$. Values for the refractive indices of water \cite{Debarre:2007kp}, glass \cite{mc2} and PNIPAM \cite{collapseEllipso} and their variation with the wavelength were obtained from the literature.
\item The density profile of the brush is modeled as a series of segments, thereby reducing the number of parameters to 6 for a two-layer profile, and 3 for a one-layer profile, as described on Fig.~\ref{fig:spectra2D}. Our rationale in this study is to identify density profiles that allow for a quantitative description of the absolute spectral reflectance while exhibiting the lowest possible level of complexity ({\it i.e.} the smallest possible number of fitting parameters). 
Our choice of a piecewise density profile containing up to 4 linear segments is motivated by the fact that it can describe or approximate many different shapes, from the simple step-like profile to more or less gradual density decays of concave or convex shapes (see SI and Fig. \ref{fig:otherprofiles}). Such a simplified model is adequate to compute the diffuse reflectance of a continuous medium such as the PNIPAM brush since the reflectance is mostly sensitive to the refractive index of the sample and its derivative, and only weakly to its second derivative (see discussion below and Fig. \ref{fig:comparisonapproximations}). At the same time, it captures the essential features of the axial density profile of the sample.
\item The total amount of polymer is conserved, which reduces further the number of free parameters to 5 (resp. 2) for a two- (resp. one-) layer model.
\end{itemize}

Under these assumptions, an analytical equation for the reflectance as a function of the 5 (2) parameters can be derived, which allows direct fitting of the reflectance curves with good accuracy. Fitting was performed using Mathematica, with the following equations for a two-phase profile: 
\begin{widetext}
\begin{equation}
\label{equreflgen}
R(\lambda)=\lvert r_0+r_1+r_2\rvert^2,
\end{equation}
with:
\begin{flalign}
&r_0 =\frac{n_{glass}-n_1}{n_{glass}+n_1} \label{equrefl1}&\\
&r_1=\frac{1}{4}\exp{\left[\frac{4i\pi n_1}{\lambda}(h_1+d_1\frac{n_2}{2(n_1-n_2)})\right]}\times\left(E_i(\frac{-2i\pi d_1n_2^2}{(n_1-n_2)\lambda})-E_i(\frac{-2i\pi d_1n_1^2}{(n_1-n_2)\lambda})\right) \label{equrefl2}&\\
&r_2=\frac{1}{4}\exp{\left[\frac{4i\pi}{\lambda}(n_2(\frac{h_s-h_1*\Phi_1}{\Phi_2})+n_1h_1+\frac{d_2n_wn_2}{2(n_2-n_w)})\right]}\times\left(E_i(\frac{-2i\pi d_2n_w^2}{(n_2-n_w)\lambda})-E_i(\frac{-2i\pi d_2n_2^2}{(n_2-n_w)\lambda})\right) \label{equrefl3}&
\end{flalign}
\end{widetext}
where $E_i$ is the exponential integral function, $\Delta n=n_{NIPAM}-n_{water}$, $n_w=n_{water}$, $n_1=\Delta n*\Phi_1+n_w$, and $n_2=\Delta n*\Phi_2+n_w$ . The five adjusted parameters $h_1$, $\Phi_1$, $\Phi_2$, $d_1$ and $d_2$ are shown on Fig. 7, and $h_s$ is the dry thickness of the PNIPAM brush measured by ellipsometry.

The formula for a single phase profile is straightforwardly obtained by setting $r_2=0$, $\Phi_2=0$ and $h_1=h_s/\Phi_1$. 

Fitting was achieved iteratively, starting from the collapsed (high temperature) spectrum for which initial values of the fit parameters can be derived from the brush dry thickness. The robustness of the solution obtained with this procedure was checked by using other initial values and, in the case of a one-phase profile, by systematically exploring the parameter space to ensure that the fit algorithm was not trapped in a local minimum. 

Fitting of spectral reflectances with the above formula relies on several assumptions, in particular:
\begin{itemize}
\item The reflection coefficients of each layer are small enough to consider that the propagation of light through the sample is not affected by the reflected wave (Born approximation). We verified this point by comparing reflectance values derived from Eq.~\ref{equreflgen} to the exact numerical calculation (see Fig. \ref{fig:comparisonapproximations}).
\item Incident light is assumed to illuminate the sample at a strictly normal incidence, whereas the illumination numerical aperture is in reality 0.15. For each wavelength $\lambda$, integrating over the illumination cone is equivalent to integrating over a range of wavelength $\left[\lambda,\lambda/\cos{\theta_{max}}\right]$, where $\theta_{max}$ is the maximum angle of incidence. Here $1/\cos{\theta_{max}}\approx1.01$ so that the reflectance should be integrated over 5-10nm. In the case of thin brushes as investigated here, the spectral variations of the reflectance have a low frequency, typically of the order of $\approx\lambda/(4 h_1 n_s)\approx100$nm (see e.g. Fig.~\ref{fig:spectrawithfits}). Integrating such curves over 5-10nm thus does not induce significant changes.
\item The density profile of the brushes is assumed to be accurately described as a series of straight lines, whereas real profiles certainly exhibit smoother features. Spectral reflectance, however, is mostly sensitive to the refractive index of the sample and its first derivative, and only indirectly to its second-order derivative (as evidence in Eq. 4). Indeed, comparison between the reflectance calculated from smoother profiles (erf functions) do not show significant differences with our simplified model (see Fig. \ref{fig:comparisonapproximations}).
Moreover, we have compared the best fits obtained with our model and with other analytical forms for the density profile, namely the parabolic and exponential decays that are frequently used to describe brushes. Such a comparison is illustrated in Fig. \ref{fig:otherprofiles} of the SI. It shows the following points: (i) our simplified model allows for excellent quantitative data fits while the two other profile shapes either miss (parabola) or only partially (exponential) reproduce the features of the reflectivity spectra, (ii) the concavity of the density profile plays an important role in capturing the proper wavelength dependence of the spectra, in particular in the swollen state of the brushes, (iii) our simplified piecewise-linear model can be used to approximate more complex density profiles, while retaining the important features of the reflectivity spectra (see Fig. \ref{fig:otherprofiles} and Fig. \ref{fig:comparisonapproximations} for the comparisons between parabolic and erf-like shapes and their discretized approximations). This supports the fact that the density profiles obtained with our simplified model are quantitatively reliable approximations of the actual density decays of the brushes, and that the quality of the fits would be only marginally enhanced by employing a higher number of segments to refine profile discretization, at the cost of a larger number of free parameters.
\item In order to ensure that the set of parameters that yields a quantitative fit of the data is unique, we do compare systematically the quality of the fits obtained with one- and two-phase models. 
Far above the LCST, when the brush is fully collapsed, the two models, whose respective parameters are all left free to vary, converge to the same best-fit solution that consists of a single step-like profile whose extension is in quantitative agreement with our independent ellipsometric measurement. This shows the uniqueness of the solution in the collapsed state, since modeling with 2 or 5 free parameters leads to the same profile. 
The same strategy is applied at all temperatures: we systematically compare the fits obtained assuming a single-phase and a two-phase model, and show that, as discussed in the manuscript, a two-phase model is mandatory in order to describe quantitatively the experimental data below and around the LCST. Once this has been identified, we further check, at selected temperatures below and at the LCST, that the profiles describing our data are indeed unique: this is done by calculating the reflectance spectrum with parameters spanning the whole range of values accessible under the physical constraints (volume fractions bounded in the range 0-1 and overall conservation of the monomer amount). We thus verify, by exploring the full parameter space, that there is only one profile that describes quantitatively the experimental data.
\end{itemize}

\bibliographystyle{achemso}
\bibliography{references,referencesSI}

\providecommand{\latin}[1]{#1}
\providecommand*\mcitethebibliography{\thebibliography}
\csname @ifundefined\endcsname{endmcitethebibliography}
  {\let\endmcitethebibliography\endthebibliography}{}
\begin{mcitethebibliography}{73}
\providecommand*\natexlab[1]{#1}
\providecommand*\mciteSetBstSublistMode[1]{}
\providecommand*\mciteSetBstMaxWidthForm[2]{}
\providecommand*\mciteBstWouldAddEndPuncttrue
  {\def\EndOfBibitem{\unskip.}}
\providecommand*\mciteBstWouldAddEndPunctfalse
  {\let\EndOfBibitem\relax}
\providecommand*\mciteSetBstMidEndSepPunct[3]{}
\providecommand*\mciteSetBstSublistLabelBeginEnd[3]{}
\providecommand*\EndOfBibitem{}
\mciteSetBstSublistMode{f}
\mciteSetBstMaxWidthForm{subitem}{(\alph{mcitesubitemcount})}
\mciteSetBstSublistLabelBeginEnd
  {\mcitemaxwidthsubitemform\space}
  {\relax}
  {\relax}

\bibitem[Azzaroni({2012})]{revBrush1}
Azzaroni,~O. \emph{{J. Polym. Sci., Part A: Polym. Chem.}} \textbf{{2012}},
  \emph{{50}}, {3225--3258}\relax
\mciteBstWouldAddEndPuncttrue
\mciteSetBstMidEndSepPunct{\mcitedefaultmidpunct}
{\mcitedefaultendpunct}{\mcitedefaultseppunct}\relax
\EndOfBibitem
\bibitem[Olivier \latin{et~al.}({2012})Olivier, Meyer, Raquez, Damman, and
  Dubois]{revBrush2}
Olivier,~A.; Meyer,~F.; Raquez,~J.-M.; Damman,~P.; Dubois,~P. \emph{{Prog.
  Polym. Sci.}} \textbf{{2012}}, \emph{{37}}, {157--181}\relax
\mciteBstWouldAddEndPuncttrue
\mciteSetBstMidEndSepPunct{\mcitedefaultmidpunct}
{\mcitedefaultendpunct}{\mcitedefaultseppunct}\relax
\EndOfBibitem
\bibitem[Barbey \latin{et~al.}({2009})Barbey, Lavanant, Paripovic, Schuewer,
  Sugnaux, Tugulu, and Klok]{revBrushAppli2}
Barbey,~R.; Lavanant,~L.; Paripovic,~D.; Schuewer,~N.; Sugnaux,~C.; Tugulu,~S.;
  Klok,~H.-A. \emph{{Chem. Rev.}} \textbf{{2009}}, \emph{{109}},
  {5437--5527}\relax
\mciteBstWouldAddEndPuncttrue
\mciteSetBstMidEndSepPunct{\mcitedefaultmidpunct}
{\mcitedefaultendpunct}{\mcitedefaultseppunct}\relax
\EndOfBibitem
\bibitem[Chen \latin{et~al.}({2010})Chen, Ferris, Zhang, Ducker, and
  Zauscher]{revBrushAppli1}
Chen,~T.; Ferris,~R.; Zhang,~J.; Ducker,~R.; Zauscher,~S. \emph{{Prog. Polym.
  Sci.}} \textbf{{2010}}, \emph{{35}}, {94--112}\relax
\mciteBstWouldAddEndPuncttrue
\mciteSetBstMidEndSepPunct{\mcitedefaultmidpunct}
{\mcitedefaultendpunct}{\mcitedefaultseppunct}\relax
\EndOfBibitem
\bibitem[Minko({2006})]{revResponsiveBrush}
Minko,~S. \emph{{Polym. Rev.}} \textbf{{2006}}, \emph{{46}}, {397--420}\relax
\mciteBstWouldAddEndPuncttrue
\mciteSetBstMidEndSepPunct{\mcitedefaultmidpunct}
{\mcitedefaultendpunct}{\mcitedefaultseppunct}\relax
\EndOfBibitem
\bibitem[Krishnamoorthy \latin{et~al.}({2014})Krishnamoorthy, Hakobyan,
  Ramstedt, and Gautrot]{revBio1}
Krishnamoorthy,~M.; Hakobyan,~S.; Ramstedt,~M.; Gautrot,~J.~E. \emph{{Chem.
  Rev.}} \textbf{{2014}}, \emph{{114}}, {10976--11026}\relax
\mciteBstWouldAddEndPuncttrue
\mciteSetBstMidEndSepPunct{\mcitedefaultmidpunct}
{\mcitedefaultendpunct}{\mcitedefaultseppunct}\relax
\EndOfBibitem
\bibitem[Cole \latin{et~al.}({2009})Cole, Voelcker, Thissen, and
  Griesser]{revBio3}
Cole,~M.~A.; Voelcker,~N.~H.; Thissen,~H.; Griesser,~H.~J.
  \emph{{Biomaterials}} \textbf{{2009}}, \emph{{30}}, {1827--1850}\relax
\mciteBstWouldAddEndPuncttrue
\mciteSetBstMidEndSepPunct{\mcitedefaultmidpunct}
{\mcitedefaultendpunct}{\mcitedefaultseppunct}\relax
\EndOfBibitem
\bibitem[Okano \latin{et~al.}({1993})Okano, Yamada, Sakai, and
  Sakurai]{cellsheet}
Okano,~T.; Yamada,~N.; Sakai,~H.; Sakurai,~Y. \emph{{J. Biomed. Mater. Res.}}
  \textbf{{1993}}, \emph{{27}}, {1243--1251}\relax
\mciteBstWouldAddEndPuncttrue
\mciteSetBstMidEndSepPunct{\mcitedefaultmidpunct}
{\mcitedefaultendpunct}{\mcitedefaultseppunct}\relax
\EndOfBibitem
\bibitem[Huber \latin{et~al.}({2003})Huber, Manginell, Samara, Kim, and
  Bunker]{protScience}
Huber,~D.; Manginell,~R.; Samara,~M.; Kim,~B.; Bunker,~B. \emph{{Science}}
  \textbf{{2003}}, \emph{{301}}, {352--354}\relax
\mciteBstWouldAddEndPuncttrue
\mciteSetBstMidEndSepPunct{\mcitedefaultmidpunct}
{\mcitedefaultendpunct}{\mcitedefaultseppunct}\relax
\EndOfBibitem
\bibitem[Miyahara \latin{et~al.}({2006})Miyahara, Nagaya, Kataoka, Yanagawa,
  Tanaka, Hao, Ishino, Ishida, Shimizu, Kangawa, Sano, Okano, Kitamura, and
  Mori]{tissue}
Miyahara,~Y.; Nagaya,~N.; Kataoka,~M.; Yanagawa,~B.; Tanaka,~K.; Hao,~H.;
  Ishino,~K.; Ishida,~H.; Shimizu,~T.; Kangawa,~K.; Sano,~S.; Okano,~T.;
  Kitamura,~S.; Mori,~H. \emph{{Nat. Med.}} \textbf{{2006}}, \emph{{12}},
  {459--465}\relax
\mciteBstWouldAddEndPuncttrue
\mciteSetBstMidEndSepPunct{\mcitedefaultmidpunct}
{\mcitedefaultendpunct}{\mcitedefaultseppunct}\relax
\EndOfBibitem
\bibitem[Kikuchi and Okano({2002})Kikuchi, and Okano]{chromato}
Kikuchi,~A.; Okano,~T. \emph{{Prog. Polym. Sci.}} \textbf{{2002}}, \emph{{27}},
  {1165--1193}\relax
\mciteBstWouldAddEndPuncttrue
\mciteSetBstMidEndSepPunct{\mcitedefaultmidpunct}
{\mcitedefaultendpunct}{\mcitedefaultseppunct}\relax
\EndOfBibitem
\bibitem[Halperin and Kroeger({2011})Halperin, and Kroeger]{Avi2}
Halperin,~A.; Kroeger,~M. \emph{{Macromolecules}} \textbf{{2011}}, \emph{{44}},
  {6986--7005}\relax
\mciteBstWouldAddEndPuncttrue
\mciteSetBstMidEndSepPunct{\mcitedefaultmidpunct}
{\mcitedefaultendpunct}{\mcitedefaultseppunct}\relax
\EndOfBibitem
\bibitem[Xue \latin{et~al.}({2012})Xue, Choi, Choi, Braun, and
  Leckband]{LeckbandProt1}
Xue,~C.; Choi,~B.-C.; Choi,~S.; Braun,~P.~V.; Leckband,~D.~E. \emph{{Adv.
  Funct. Mater.}} \textbf{{2012}}, \emph{{22}}, {2394--2401}\relax
\mciteBstWouldAddEndPuncttrue
\mciteSetBstMidEndSepPunct{\mcitedefaultmidpunct}
{\mcitedefaultendpunct}{\mcitedefaultseppunct}\relax
\EndOfBibitem
\bibitem[Halperin and Kroeger({2012})Halperin, and Kroeger]{Avi3}
Halperin,~A.; Kroeger,~M. \emph{{Biomaterials}} \textbf{{2012}}, \emph{{33}},
  {4975--4987}\relax
\mciteBstWouldAddEndPuncttrue
\mciteSetBstMidEndSepPunct{\mcitedefaultmidpunct}
{\mcitedefaultendpunct}{\mcitedefaultseppunct}\relax
\EndOfBibitem
\bibitem[Wang \latin{et~al.}({1998})Wang, Qiu, and Wu]{Wu2}
Wang,~X.; Qiu,~X.; Wu,~C. \emph{{Macromolecules}} \textbf{{1998}}, \emph{{31}},
  {2972--2976}\relax
\mciteBstWouldAddEndPuncttrue
\mciteSetBstMidEndSepPunct{\mcitedefaultmidpunct}
{\mcitedefaultendpunct}{\mcitedefaultseppunct}\relax
\EndOfBibitem
\bibitem[Balamurugan \latin{et~al.}({2003})Balamurugan, Mendez, Balamurugan,
  O'Brien, and Lopez]{SPR}
Balamurugan,~S.; Mendez,~S.; Balamurugan,~S.; O'Brien,~M.; Lopez,~G.
  \emph{{Langmuir}} \textbf{{2003}}, \emph{{19}}, {2545--2549}\relax
\mciteBstWouldAddEndPuncttrue
\mciteSetBstMidEndSepPunct{\mcitedefaultmidpunct}
{\mcitedefaultendpunct}{\mcitedefaultseppunct}\relax
\EndOfBibitem
\bibitem[Zhang({2004})]{QCM1}
Zhang,~G. \emph{{Macromolecules}} \textbf{{2004}}, \emph{{37}},
  {6553--6557}\relax
\mciteBstWouldAddEndPuncttrue
\mciteSetBstMidEndSepPunct{\mcitedefaultmidpunct}
{\mcitedefaultendpunct}{\mcitedefaultseppunct}\relax
\EndOfBibitem
\bibitem[Annaka \latin{et~al.}({2007})Annaka, Yahiro, Nagase, Kikuchi, and
  Okano]{QCM2}
Annaka,~M.; Yahiro,~C.; Nagase,~K.; Kikuchi,~A.; Okano,~T. \emph{{Polymer}}
  \textbf{{2007}}, \emph{{48}}, {5713--5720}\relax
\mciteBstWouldAddEndPuncttrue
\mciteSetBstMidEndSepPunct{\mcitedefaultmidpunct}
{\mcitedefaultendpunct}{\mcitedefaultseppunct}\relax
\EndOfBibitem
\bibitem[Zhang and Wu({2009})Zhang, and Wu]{QCM3}
Zhang,~G.; Wu,~C. \emph{{Macromol. Rapid Commun.}} \textbf{{2009}},
  \emph{{30}}, {328--335}\relax
\mciteBstWouldAddEndPuncttrue
\mciteSetBstMidEndSepPunct{\mcitedefaultmidpunct}
{\mcitedefaultendpunct}{\mcitedefaultseppunct}\relax
\EndOfBibitem
\bibitem[Liu and Zhang({2005})Liu, and Zhang]{QCM4}
Liu,~G.; Zhang,~G. \emph{{J. Phys. Chem. B}} \textbf{{2005}}, \emph{{109}},
  {743--747}\relax
\mciteBstWouldAddEndPuncttrue
\mciteSetBstMidEndSepPunct{\mcitedefaultmidpunct}
{\mcitedefaultendpunct}{\mcitedefaultseppunct}\relax
\EndOfBibitem
\bibitem[Ishida and Biggs({2007})Ishida, and Biggs]{AFM-QCM1}
Ishida,~N.; Biggs,~S. \emph{{Langmuir}} \textbf{{2007}}, \emph{{23}},
  {11083--11088}\relax
\mciteBstWouldAddEndPuncttrue
\mciteSetBstMidEndSepPunct{\mcitedefaultmidpunct}
{\mcitedefaultendpunct}{\mcitedefaultseppunct}\relax
\EndOfBibitem
\bibitem[Ishida and Biggs({2010})Ishida, and Biggs]{AFM-QCM2}
Ishida,~N.; Biggs,~S. \emph{{Macromolecules}} \textbf{{2010}}, \emph{{43}},
  {7269--7276}\relax
\mciteBstWouldAddEndPuncttrue
\mciteSetBstMidEndSepPunct{\mcitedefaultmidpunct}
{\mcitedefaultendpunct}{\mcitedefaultseppunct}\relax
\EndOfBibitem
\bibitem[Plunkett \latin{et~al.}({2006})Plunkett, Zhu, Moore, and
  Leckband]{LeckbandSFA}
Plunkett,~K.; Zhu,~X.; Moore,~J.; Leckband,~D. \emph{{Langmuir}}
  \textbf{{2006}}, \emph{{22}}, {4259--4266}\relax
\mciteBstWouldAddEndPuncttrue
\mciteSetBstMidEndSepPunct{\mcitedefaultmidpunct}
{\mcitedefaultendpunct}{\mcitedefaultseppunct}\relax
\EndOfBibitem
\bibitem[Malham and Bureau({2010})Malham, and Bureau]{bureau1}
Malham,~I.~B.; Bureau,~L. \emph{{Langmuir}} \textbf{{2010}}, \emph{{26}},
  {4762--4768}\relax
\mciteBstWouldAddEndPuncttrue
\mciteSetBstMidEndSepPunct{\mcitedefaultmidpunct}
{\mcitedefaultendpunct}{\mcitedefaultseppunct}\relax
\EndOfBibitem
\bibitem[Kidoaki \latin{et~al.}({2001})Kidoaki, Ohya, Nakayama, and
  Matsuda]{AFM1}
Kidoaki,~S.; Ohya,~S.; Nakayama,~Y.; Matsuda,~T. \emph{{Langmuir}}
  \textbf{{2001}}, \emph{{17}}, {2402--2407}\relax
\mciteBstWouldAddEndPuncttrue
\mciteSetBstMidEndSepPunct{\mcitedefaultmidpunct}
{\mcitedefaultendpunct}{\mcitedefaultseppunct}\relax
\EndOfBibitem
\bibitem[Ishida and Kobayashi({2006})Ishida, and Kobayashi]{AFM2}
Ishida,~N.; Kobayashi,~M. \emph{{J. Colloid Interface Sci.}} \textbf{{2006}},
  \emph{{297}}, {513--519}\relax
\mciteBstWouldAddEndPuncttrue
\mciteSetBstMidEndSepPunct{\mcitedefaultmidpunct}
{\mcitedefaultendpunct}{\mcitedefaultseppunct}\relax
\EndOfBibitem
\bibitem[Benetti \latin{et~al.}({2007})Benetti, Zapotoczny, and Vancso]{AFM3}
Benetti,~E.~M.; Zapotoczny,~S.; Vancso,~G.~J. \emph{{Adv. Mat.}}
  \textbf{{2007}}, \emph{{19}}, {268--271}\relax
\mciteBstWouldAddEndPuncttrue
\mciteSetBstMidEndSepPunct{\mcitedefaultmidpunct}
{\mcitedefaultendpunct}{\mcitedefaultseppunct}\relax
\EndOfBibitem
\bibitem[Yim \latin{et~al.}({2004})Yim, Kent, Mendez, Balamurugan, Balamurugan,
  Lopez, and Satija]{Neutron1}
Yim,~H.; Kent,~M.; Mendez,~S.; Balamurugan,~S.; Balamurugan,~S.; Lopez,~G.;
  Satija,~S. \emph{{Macromolecules}} \textbf{{2004}}, \emph{{37}},
  {1994--1997}\relax
\mciteBstWouldAddEndPuncttrue
\mciteSetBstMidEndSepPunct{\mcitedefaultmidpunct}
{\mcitedefaultendpunct}{\mcitedefaultseppunct}\relax
\EndOfBibitem
\bibitem[Yim \latin{et~al.}({2006})Yim, Kent, Mendez, Lopez, Satija, and
  Seo]{Neutron2}
Yim,~H.; Kent,~M.; Mendez,~S.; Lopez,~G.; Satija,~S.; Seo,~Y.
  \emph{{Macromolecules}} \textbf{{2006}}, \emph{{39}}, {3420--3426}\relax
\mciteBstWouldAddEndPuncttrue
\mciteSetBstMidEndSepPunct{\mcitedefaultmidpunct}
{\mcitedefaultendpunct}{\mcitedefaultseppunct}\relax
\EndOfBibitem
\bibitem[Yim \latin{et~al.}({2005})Yim, Kent, Satija, Mendez, Balamurugan,
  Balamurugan, and Lopez]{Neutron3}
Yim,~H.; Kent,~M.; Satija,~S.; Mendez,~S.; Balamurugan,~S.; Balamurugan,~S.;
  Lopez,~G. \emph{{Phys. Rev. E}} \textbf{{2005}}, \emph{{72}}\relax
\mciteBstWouldAddEndPuncttrue
\mciteSetBstMidEndSepPunct{\mcitedefaultmidpunct}
{\mcitedefaultendpunct}{\mcitedefaultseppunct}\relax
\EndOfBibitem
\bibitem[Kooij \latin{et~al.}({2012})Kooij, Sui, Hempenius, Zandvliet, and
  Vancso]{collapseEllipso}
Kooij,~E.~S.; Sui,~X.; Hempenius,~M.~A.; Zandvliet,~H. J.~W.; Vancso,~G.~J.
  \emph{{J. Phys. Chem. B}} \textbf{{2012}}, \emph{{116}}, {9261--9268}\relax
\mciteBstWouldAddEndPuncttrue
\mciteSetBstMidEndSepPunct{\mcitedefaultmidpunct}
{\mcitedefaultendpunct}{\mcitedefaultseppunct}\relax
\EndOfBibitem
\bibitem[Bittrich \latin{et~al.}({2012})Bittrich, Burkert, M{\"u}ller,
  Eichhorn, Stamm, and Uhlmann]{Ellipso2}
Bittrich,~E.; Burkert,~S.; M{\"u}ller,~M.; Eichhorn,~K.-J.; Stamm,~M.;
  Uhlmann,~P. \emph{{Langmuir}} \textbf{{2012}}, \emph{{28}},
  {3439--3448}\relax
\mciteBstWouldAddEndPuncttrue
\mciteSetBstMidEndSepPunct{\mcitedefaultmidpunct}
{\mcitedefaultendpunct}{\mcitedefaultseppunct}\relax
\EndOfBibitem
\bibitem[Biesalski \latin{et~al.}({1999})Biesalski, R{\"u}he, and
  Johannsmann]{Ellipso3}
Biesalski,~M.; R{\"u}he,~J.; Johannsmann,~D. \emph{{J. Chem. Phys.}}
  \textbf{{1999}}, \emph{{111}}, {7029--7037}\relax
\mciteBstWouldAddEndPuncttrue
\mciteSetBstMidEndSepPunct{\mcitedefaultmidpunct}
{\mcitedefaultendpunct}{\mcitedefaultseppunct}\relax
\EndOfBibitem
\bibitem[Gast(1992)]{gast}
Gast,~A.~P. In \emph{Physics of polymer surfaces and interfaces}; Sanchez,~I.,
  Fitzpatrick,~L., Eds.; Butterworth-Heinemann, 1992; Chapter 11, pp
  245--262\relax
\mciteBstWouldAddEndPuncttrue
\mciteSetBstMidEndSepPunct{\mcitedefaultmidpunct}
{\mcitedefaultendpunct}{\mcitedefaultseppunct}\relax
\EndOfBibitem
\bibitem[Halperin \latin{et~al.}({2015})Halperin, Kroeger, and Winnik]{Winnik}
Halperin,~A.; Kroeger,~M.; Winnik,~F.~M. \emph{{Angew. Chem. Int. Ed.}}
  \textbf{{2015}}, \emph{{54}}, {15342--15367}\relax
\mciteBstWouldAddEndPuncttrue
\mciteSetBstMidEndSepPunct{\mcitedefaultmidpunct}
{\mcitedefaultendpunct}{\mcitedefaultseppunct}\relax
\EndOfBibitem
\bibitem[Okada \latin{et~al.}({2015})Okada, Akiyama, Kobayashi, Ninomiya,
  Kanazawa, Yamato, and Okano]{refracto}
Okada,~F.; Akiyama,~Y.; Kobayashi,~J.; Ninomiya,~H.; Kanazawa,~H.; Yamato,~M.;
  Okano,~T. \emph{{J. Nanopart. Res.}} \textbf{{2015}}, \emph{{17}},
  {148}\relax
\mciteBstWouldAddEndPuncttrue
\mciteSetBstMidEndSepPunct{\mcitedefaultmidpunct}
{\mcitedefaultendpunct}{\mcitedefaultseppunct}\relax
\EndOfBibitem
\bibitem[Afroze \latin{et~al.}({2000})Afroze, Nies, and Berghmans]{Afroze}
Afroze,~F.; Nies,~E.; Berghmans,~H. \emph{{J. Mol. Struct.}} \textbf{{2000}},
  \emph{{554}}, {55--68}\relax
\mciteBstWouldAddEndPuncttrue
\mciteSetBstMidEndSepPunct{\mcitedefaultmidpunct}
{\mcitedefaultendpunct}{\mcitedefaultseppunct}\relax
\EndOfBibitem
\bibitem[Baulin and Halperin({2002})Baulin, and Halperin]{2state}
Baulin,~V.~A.; Halperin,~A. \emph{{Macromolecules}} \textbf{{2002}},
  \emph{{35}}, {6432--6438}\relax
\mciteBstWouldAddEndPuncttrue
\mciteSetBstMidEndSepPunct{\mcitedefaultmidpunct}
{\mcitedefaultendpunct}{\mcitedefaultseppunct}\relax
\EndOfBibitem
\bibitem[Baulin and Halperin({2003})Baulin, and Halperin]{Avi4}
Baulin,~V.; Halperin,~A. \emph{{Macromol. Theory Simul.}} \textbf{{2003}},
  \emph{{12}}, {549--559}\relax
\mciteBstWouldAddEndPuncttrue
\mciteSetBstMidEndSepPunct{\mcitedefaultmidpunct}
{\mcitedefaultendpunct}{\mcitedefaultseppunct}\relax
\EndOfBibitem
\bibitem[Wagner \latin{et~al.}({1993})Wagner, Brochard-Wyart, Hervet, and
  De~Gennes]{Ncluster}
Wagner,~M.; Brochard-Wyart,~F.; Hervet,~H.; De~Gennes,~P. \emph{{Colloid.
  Polym. Sci.}} \textbf{{1993}}, \emph{{271}}, {621--628}\relax
\mciteBstWouldAddEndPuncttrue
\mciteSetBstMidEndSepPunct{\mcitedefaultmidpunct}
{\mcitedefaultendpunct}{\mcitedefaultseppunct}\relax
\EndOfBibitem
\bibitem[Baulin \latin{et~al.}({2003})Baulin, Zhulina, and Halperin]{Avi1}
Baulin,~V.; Zhulina,~E.; Halperin,~A. \emph{{J. Chem. Phys.}} \textbf{{2003}},
  \emph{{119}}, {10977--10988}\relax
\mciteBstWouldAddEndPuncttrue
\mciteSetBstMidEndSepPunct{\mcitedefaultmidpunct}
{\mcitedefaultendpunct}{\mcitedefaultseppunct}\relax
\EndOfBibitem
\bibitem[Xia \latin{et~al.}({2005})Xia, Yin, Burke, and St{\"o}ver]{length}
Xia,~Y.; Yin,~X.; Burke,~N. A.~D.; St{\"o}ver,~H. D.~H. \emph{{Macromolecules}}
  \textbf{{2005}}, \emph{{38}}, {5937--5943}\relax
\mciteBstWouldAddEndPuncttrue
\mciteSetBstMidEndSepPunct{\mcitedefaultmidpunct}
{\mcitedefaultendpunct}{\mcitedefaultseppunct}\relax
\EndOfBibitem
\bibitem[L{\'e}onforte and M{\"u}ller({2015})L{\'e}onforte, and
  M{\"u}ller]{leonforte}
L{\'e}onforte,~F.; M{\"u}ller,~M. \emph{{ACS Appl. Mater. Interfaces}}
  \textbf{{2015}}, \emph{{7}}, {12450--12462}\relax
\mciteBstWouldAddEndPuncttrue
\mciteSetBstMidEndSepPunct{\mcitedefaultmidpunct}
{\mcitedefaultendpunct}{\mcitedefaultseppunct}\relax
\EndOfBibitem
\bibitem[De~Gennes({1980})]{degennes}
De~Gennes,~P.~G. \emph{{Macromolecules}} \textbf{{1980}}, \emph{{13}},
  {1069--1075}\relax
\mciteBstWouldAddEndPuncttrue
\mciteSetBstMidEndSepPunct{\mcitedefaultmidpunct}
{\mcitedefaultendpunct}{\mcitedefaultseppunct}\relax
\EndOfBibitem
\bibitem[Wijmans and Zhulina({1993})Wijmans, and Zhulina]{zhulina}
Wijmans,~C.~M.; Zhulina,~E.~B. \emph{{Macromolecules}} \textbf{{1993}},
  \emph{{26}}, {7214--7224}\relax
\mciteBstWouldAddEndPuncttrue
\mciteSetBstMidEndSepPunct{\mcitedefaultmidpunct}
{\mcitedefaultendpunct}{\mcitedefaultseppunct}\relax
\EndOfBibitem
\bibitem[Yamamoto \latin{et~al.}({2000})Yamamoto, Ejaz, Tsujii, and Fukuda]{nu}
Yamamoto,~S.; Ejaz,~M.; Tsujii,~Y.; Fukuda,~T. \emph{{Macromolecules}}
  \textbf{{2000}}, \emph{{33}}, {5608--5612}\relax
\mciteBstWouldAddEndPuncttrue
\mciteSetBstMidEndSepPunct{\mcitedefaultmidpunct}
{\mcitedefaultendpunct}{\mcitedefaultseppunct}\relax
\EndOfBibitem
\bibitem[Note1()]{Note1}
We estimate $a$ from $\rho =M_0/(N_Aa^3)$, with $\rho =0.95$ g.cm$^{-3}$ the
  NIPAM monomer density, $M_0=113$ g.mol$^{-1}$, and $N_A$ the Avogadro
  constant.\relax
\mciteBstWouldAddEndPunctfalse
\mciteSetBstMidEndSepPunct{\mcitedefaultmidpunct}
{}{\mcitedefaultseppunct}\relax
\EndOfBibitem
\bibitem[Grest and Murat({1993})Grest, and Murat]{grest}
Grest,~G.~S.; Murat,~M. \emph{{Macromolecules}} \textbf{{1993}}, \emph{{26}},
  {3108--3117}\relax
\mciteBstWouldAddEndPuncttrue
\mciteSetBstMidEndSepPunct{\mcitedefaultmidpunct}
{\mcitedefaultendpunct}{\mcitedefaultseppunct}\relax
\EndOfBibitem
\bibitem[Milner \latin{et~al.}({1989})Milner, Witten, and Cates]{milner}
Milner,~S.~T.; Witten,~T.~A.; Cates,~M.~E. \emph{{Macromolecules}}
  \textbf{{1989}}, \emph{{22}}, {853--861}\relax
\mciteBstWouldAddEndPuncttrue
\mciteSetBstMidEndSepPunct{\mcitedefaultmidpunct}
{\mcitedefaultendpunct}{\mcitedefaultseppunct}\relax
\EndOfBibitem
\bibitem[Dimitrov \latin{et~al.}({2007})Dimitrov, Milchev, and Binder]{binder}
Dimitrov,~D.~I.; Milchev,~A.; Binder,~K. \emph{{J. Chem. Phys.}}
  \textbf{{2007}}, \emph{{127}}, {084905.1--084905.9}\relax
\mciteBstWouldAddEndPuncttrue
\mciteSetBstMidEndSepPunct{\mcitedefaultmidpunct}
{\mcitedefaultendpunct}{\mcitedefaultseppunct}\relax
\EndOfBibitem
\bibitem[de~Vos and Leermakers({2009})de~Vos, and Leermakers]{devos}
de~Vos,~W.~M.; Leermakers,~F. A.~M. \emph{{Polymer}} \textbf{{2009}},
  \emph{{50}}, {305--316}\relax
\mciteBstWouldAddEndPuncttrue
\mciteSetBstMidEndSepPunct{\mcitedefaultmidpunct}
{\mcitedefaultendpunct}{\mcitedefaultseppunct}\relax
\EndOfBibitem
\bibitem[Shivapooja \latin{et~al.}({2012})Shivapooja, Ista, Canavan, and
  Lopez]{ARGET1}
Shivapooja,~P.; Ista,~L.~K.; Canavan,~H.~E.; Lopez,~G.~P.
  \emph{{Biointerphases}} \textbf{{2012}}, \emph{{7}}, {32}\relax
\mciteBstWouldAddEndPuncttrue
\mciteSetBstMidEndSepPunct{\mcitedefaultmidpunct}
{\mcitedefaultendpunct}{\mcitedefaultseppunct}\relax
\EndOfBibitem
\bibitem[Patil \latin{et~al.}({2015})Patil, Turgman-Cohen, Srogl, Kiserow, and
  Genzer]{degraft}
Patil,~R.~R.; Turgman-Cohen,~S.; Srogl,~J.; Kiserow,~D.; Genzer,~J.
  \emph{{Langmuir}} \textbf{{2015}}, \emph{{31}}, {2372--2381}\relax
\mciteBstWouldAddEndPuncttrue
\mciteSetBstMidEndSepPunct{\mcitedefaultmidpunct}
{\mcitedefaultendpunct}{\mcitedefaultseppunct}\relax
\EndOfBibitem
\bibitem[Kang \latin{et~al.}({2016})Kang, Crockett, and Spencer]{degraft2}
Kang,~C.; Crockett,~R.; Spencer,~N.~D. \emph{{Polym Chem.}} \textbf{{2016}},
  \emph{{7}}, {302--309}\relax
\mciteBstWouldAddEndPuncttrue
\mciteSetBstMidEndSepPunct{\mcitedefaultmidpunct}
{\mcitedefaultendpunct}{\mcitedefaultseppunct}\relax
\EndOfBibitem
\bibitem[Gorman \latin{et~al.}({2008})Gorman, Petrie, and Genzer]{genzer2}
Gorman,~C.~B.; Petrie,~R.~J.; Genzer,~J. \emph{{Macromolecules}}
  \textbf{{2008}}, \emph{{41}}, {4856--4865}\relax
\mciteBstWouldAddEndPuncttrue
\mciteSetBstMidEndSepPunct{\mcitedefaultmidpunct}
{\mcitedefaultendpunct}{\mcitedefaultseppunct}\relax
\EndOfBibitem
\bibitem[Genzer({2006})]{genzer1}
Genzer,~J. \emph{{Macromolecules}} \textbf{{2006}}, \emph{{39}},
  {7157--7169}\relax
\mciteBstWouldAddEndPuncttrue
\mciteSetBstMidEndSepPunct{\mcitedefaultmidpunct}
{\mcitedefaultendpunct}{\mcitedefaultseppunct}\relax
\EndOfBibitem
\bibitem[Xue \latin{et~al.}({2011})Xue, Yonet-Tanyeri, Brouette, Sferrazza,
  Braun, and Leckband]{LeckbandProt2}
Xue,~C.; Yonet-Tanyeri,~N.; Brouette,~N.; Sferrazza,~M.; Braun,~P.~V.;
  Leckband,~D.~E. \emph{{Langmuir}} \textbf{{2011}}, \emph{{27}},
  {8810--8818}\relax
\mciteBstWouldAddEndPuncttrue
\mciteSetBstMidEndSepPunct{\mcitedefaultmidpunct}
{\mcitedefaultendpunct}{\mcitedefaultseppunct}\relax
\EndOfBibitem
\bibitem[Cheng \latin{et~al.}({2006})Cheng, Shen, and Wu]{Wu1}
Cheng,~H.; Shen,~L.; Wu,~C. \emph{{Macromolecules}} \textbf{{2006}},
  \emph{{39}}, {2325--2329}\relax
\mciteBstWouldAddEndPuncttrue
\mciteSetBstMidEndSepPunct{\mcitedefaultmidpunct}
{\mcitedefaultendpunct}{\mcitedefaultseppunct}\relax
\EndOfBibitem
\bibitem[Mc()]{Mc}
\url{http://polymerdatabase.com/polymers/polyn-isopropylacrylamide.html}\relax
\mciteBstWouldAddEndPuncttrue
\mciteSetBstMidEndSepPunct{\mcitedefaultmidpunct}
{\mcitedefaultendpunct}{\mcitedefaultseppunct}\relax
\EndOfBibitem
\bibitem[Yuan \latin{et~al.}({2006})Yuan, Wang, Han, and Wu]{Wu3}
Yuan,~G.; Wang,~X.; Han,~C.~C.; Wu,~C. \emph{{Macromolecules}} \textbf{{2006}},
  \emph{{39}}, {3642--3647}\relax
\mciteBstWouldAddEndPuncttrue
\mciteSetBstMidEndSepPunct{\mcitedefaultmidpunct}
{\mcitedefaultendpunct}{\mcitedefaultseppunct}\relax
\EndOfBibitem
\bibitem[Pearson and Helfand({1984})Pearson, and Helfand]{HP}
Pearson,~D.~A.; Helfand,~E. \emph{{Macromolecules}} \textbf{{1984}},
  \emph{{17}}, {888--895}\relax
\mciteBstWouldAddEndPuncttrue
\mciteSetBstMidEndSepPunct{\mcitedefaultmidpunct}
{\mcitedefaultendpunct}{\mcitedefaultseppunct}\relax
\EndOfBibitem
\bibitem[Vega \latin{et~al.}({2005})Vega, Gomez, Roth, Ressia, Villar, and
  Valles]{Vega}
Vega,~D.~A.; Gomez,~L.~R.; Roth,~L.~E.; Ressia,~J.~A.; Villar,~M.~A.;
  Valles,~E.~M. \emph{{Phys. Rev. Lett.}} \textbf{{2005}}, \emph{{95}},
  {166002.1--166002.4}\relax
\mciteBstWouldAddEndPuncttrue
\mciteSetBstMidEndSepPunct{\mcitedefaultmidpunct}
{\mcitedefaultendpunct}{\mcitedefaultseppunct}\relax
\EndOfBibitem
\bibitem[Ajdari \latin{et~al.}({1994})Ajdari, Brochard-Wyart, de~Gennes,
  Leibler, Viovy, and Rubinstein]{Ajdari}
Ajdari,~A.; Brochard-Wyart,~F.; de~Gennes,~P.; Leibler,~L.; Viovy,~J.;
  Rubinstein,~M. \emph{{Physica A}} \textbf{{1994}}, \emph{{204}},
  {17--39}\relax
\mciteBstWouldAddEndPuncttrue
\mciteSetBstMidEndSepPunct{\mcitedefaultmidpunct}
{\mcitedefaultendpunct}{\mcitedefaultseppunct}\relax
\EndOfBibitem
\bibitem[Note2()]{Note2}
We consider here that $N_e$ is approximately equal to half the critical length
  for entanglements $N_c$, as suggested from rheology studies on polymer
  melts\cite {Colby}, and use $M_c=23$ kg.mol$^{-1}$\relax
\mciteBstWouldAddEndPuncttrue
\mciteSetBstMidEndSepPunct{\mcitedefaultmidpunct}
{\mcitedefaultendpunct}{\mcitedefaultseppunct}\relax
\EndOfBibitem
\bibitem[Rubinstein and Colby(2003)Rubinstein, and Colby]{Colby}
Rubinstein,~M.; Colby,~R.~H. \emph{Polymer physics}; Oxford Univ. Press: New
  York, 2003; Chapter 9, pp 361--422\relax
\mciteBstWouldAddEndPuncttrue
\mciteSetBstMidEndSepPunct{\mcitedefaultmidpunct}
{\mcitedefaultendpunct}{\mcitedefaultseppunct}\relax
\EndOfBibitem
\bibitem[Lee \latin{et~al.}({2004})Lee, Abrams, Johner, and Obukhov]{Johner}
Lee,~N.~K.; Abrams,~C.~F.; Johner,~A.; Obukhov,~S. \emph{{Macromolecules}}
  \textbf{{2004}}, \emph{{37}}, {651--661}\relax
\mciteBstWouldAddEndPuncttrue
\mciteSetBstMidEndSepPunct{\mcitedefaultmidpunct}
{\mcitedefaultendpunct}{\mcitedefaultseppunct}\relax
\EndOfBibitem
\bibitem[Limozin and Sengupta({2009})Limozin, and Sengupta]{RICM}
Limozin,~L.; Sengupta,~K. \emph{{ChemPhysChem}} \textbf{{2009}}, \emph{{10}},
  {2752--2768}\relax
\mciteBstWouldAddEndPuncttrue
\mciteSetBstMidEndSepPunct{\mcitedefaultmidpunct}
{\mcitedefaultendpunct}{\mcitedefaultseppunct}\relax
\EndOfBibitem
\bibitem[Killeen and Breiland(1994)Killeen, and Breiland]{Killeen:1994ws}
Killeen,~K.~P.; Breiland,~W.~G. \emph{J. Electron. Mater.} \textbf{1994},
  \emph{23}, 179--183\relax
\mciteBstWouldAddEndPuncttrue
\mciteSetBstMidEndSepPunct{\mcitedefaultmidpunct}
{\mcitedefaultendpunct}{\mcitedefaultseppunct}\relax
\EndOfBibitem
\bibitem[Gauglitz \latin{et~al.}(1993)Gauglitz, Brecht, Kraus, and
  Mahm]{Gauglitz:1993tb}
Gauglitz,~G.; Brecht,~A.; Kraus,~G.; Mahm,~W. \emph{Sens. Actuators, B}
  \textbf{1993}, \emph{11}, 21--27\relax
\mciteBstWouldAddEndPuncttrue
\mciteSetBstMidEndSepPunct{\mcitedefaultmidpunct}
{\mcitedefaultendpunct}{\mcitedefaultseppunct}\relax
\EndOfBibitem
\bibitem[Debarre and Beaurepaire(2007)Debarre, and Beaurepaire]{Debarre:2007kp}
Debarre,~D.; Beaurepaire,~E. \emph{Biophysical Journal} \textbf{2007},
  \emph{92}, 603--612\relax
\mciteBstWouldAddEndPuncttrue
\mciteSetBstMidEndSepPunct{\mcitedefaultmidpunct}
{\mcitedefaultendpunct}{\mcitedefaultseppunct}\relax
\EndOfBibitem
\bibitem[mc2()]{mc2}
\url{http://refractiveindex.info/?shelf=glass&book=BK7&page=SCHOTT}\relax
\mciteBstWouldAddEndPuncttrue
\mciteSetBstMidEndSepPunct{\mcitedefaultmidpunct}
{\mcitedefaultendpunct}{\mcitedefaultseppunct}\relax
\EndOfBibitem
\bibitem[Brown \latin{et~al.}({2005})Brown, Khan, Steinbock, and
  Huck]{huckATRP}
Brown,~A.; Khan,~N.; Steinbock,~L.; Huck,~W. \emph{{Eur. Polym. J.}}
  \textbf{{2005}}, \emph{{41}}, {1757--1765}\relax
\mciteBstWouldAddEndPuncttrue
\mciteSetBstMidEndSepPunct{\mcitedefaultmidpunct}
{\mcitedefaultendpunct}{\mcitedefaultseppunct}\relax
\EndOfBibitem
\bibitem[Matyjaszewski \latin{et~al.}({2007})Matyjaszewski, Dong, Jakubowski,
  Pietrasik, and Kusumo]{ARGET2}
Matyjaszewski,~K.; Dong,~H.; Jakubowski,~W.; Pietrasik,~J.; Kusumo,~A.
  \emph{{Langmuir}} \textbf{{2007}}, \emph{{23}}, {4528--4531}\relax
\mciteBstWouldAddEndPuncttrue
\mciteSetBstMidEndSepPunct{\mcitedefaultmidpunct}
{\mcitedefaultendpunct}{\mcitedefaultseppunct}\relax
\EndOfBibitem
\end{mcitethebibliography}


\onecolumngrid

\vspace{5cm}

\section*{Supplementary Information}

\subsection*{Materials}
\label{subsec:chemicals}

N-isopropylacrylamide (NIPAM, 99\%), 3-Aminopropyltriethoxysilane (APTES, 99\%), triethylamine (TEA, 99.7\% pure), copper (II) bromide (CuBr$_2$, 99\% extra pure), 1,1,7,7-pentamethyldiethylenetriamine (PMDETA, 99\%), 2-bromo-2-methylpropionyl bromide (BMPB, 98\% pure) and propionyl bromide (PB, 95\%) were purchased from Acros Organics. Ascorbic acid (AA, 99\%) was obtained from Sigma Aldrich. All reagents were used as received except NIPAM, which was recrystallized twice in n-hexane (Normapur, VWR) and then dried before use. Absolute ethanol and dichloromethane (DCM) were obtained from Fischer Chemicals (Laboratory Grade).
Ultrapure water (18.2 M$\Omega$.cm) was obtained from a Millipore Synergy system.

Glass coverslips (N$^{\circ}$1 type, thickness $\sim$150 $\mu$m, Thermo Scientific Menzel) of 20 mm diameter and silicon wafers (2'' diameter, 100 orientation, ACM France) were used as the substrates. Si wafers were diced into 1 cm$^2$ pieces before use.

Optical quality muscovite mica was obtained from JBG Metafix (France) under the form of 10$\times$10 cm$^2$ and 0.1 mm-thick plates.

\subsection*{Preparation of the samples}
\label{subsec:samples}

Glass and silicon substrates were first thoroughly rinsed with absolute ethanol and water, and blown dry in a stream of nitrogen. They were then transferred to a plasma cleaner (Femto, Diener Electronics Germany, operated at 80W), where they were exposed for 6 minutes to a plasma generated in water vapor, at a pressure of 0.4 mbar. Such a treatment has been shown to yield clean and highly hydroxylated glass or oxide surfaces \cite{bureau1}.  

\begin{figure}[h]
$$
\includegraphics[width=8cm]{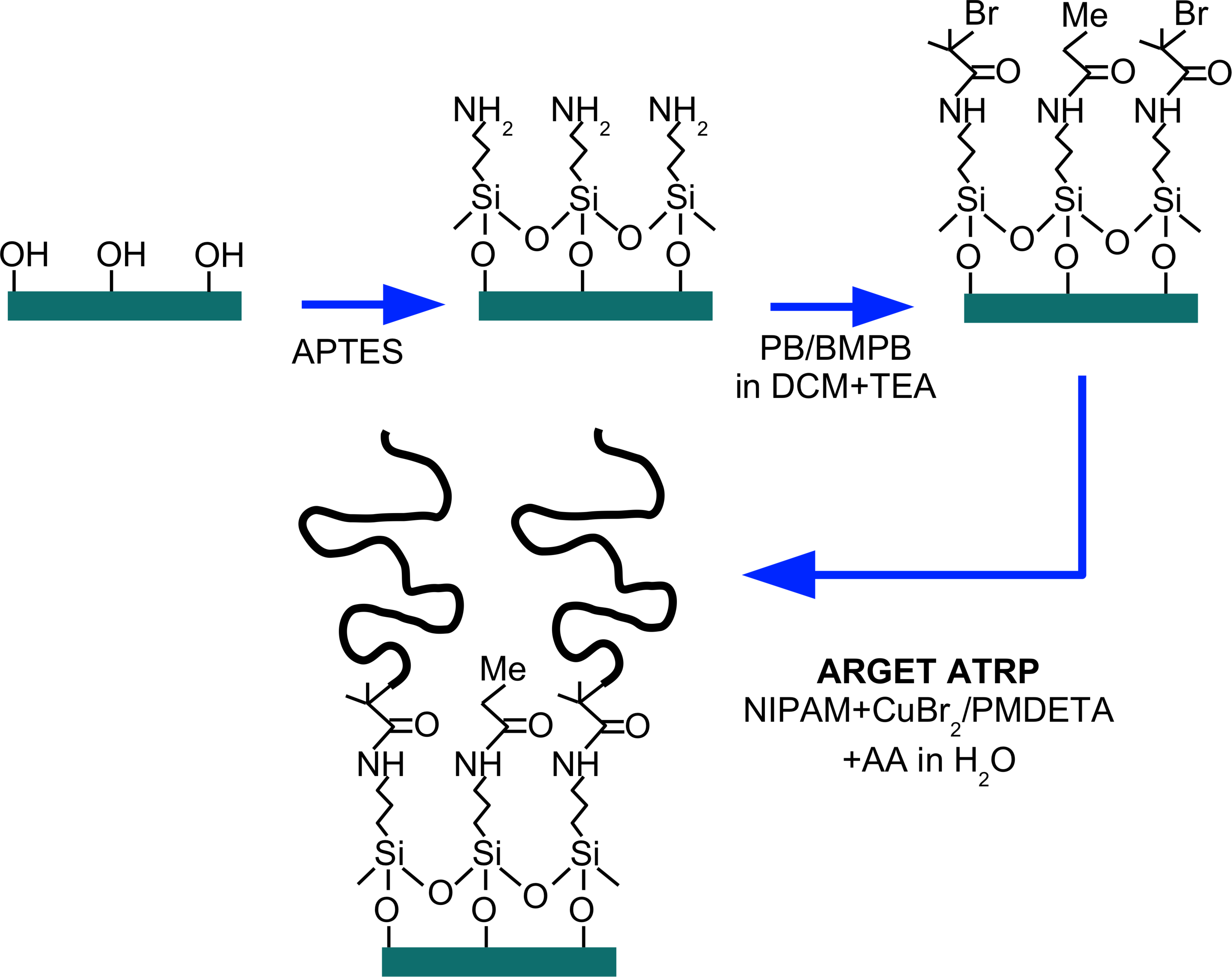}
$$
\caption{\label{fig:scheme1}Scheme of surface functionalization protocol: starting from a hydroxilated surface, APTES is grafted and derivitized in order to obtain a layer of surface-bound ATRP initiators, from which PNIPAM polymerization is performed.}
\end{figure}

Substrates were then functionalized via a surface-initiated Atom Transfer Radical Polymerization (ATRP) procedure, following the protocol summarized in Fig. \ref{fig:scheme1}, which was adapted from previous works \cite{bureau1,huckATRP,ARGET1}. 

\subsubsection*{Grafting of ATRP initiator}
\label{subsubsec:initiator}

A volume of 50 mL of a solution of APTES in water (c$_{APTES}$=8$\times$10$^{-3}$ M) was prepared and stirred for two hours before use, in order to per-hydrolyze the APTES ethoxysilane groups. The solution was then passed through a 0.2 $\mu$m syringe filter in order to remove possible silane aggregates, and immediately used. The plasma-cleaned substrates were immersed in this solution for 1 minute, then rinsed copiously with water and ethanol, and carefully dried in a nitrogen stream. 

The -NH$_2$ groups of the deposited APTES layer were then derivatized as follows. 

The maximum initiator density investigated in this study was obtained by immersing the APTES-treated surfaces for 1 minute in a solution of BMPB (0.25 mL) and TEA (1.2 mL) in DCM (25 mL), followed by rinsing in DCM, ethanol and water. This leads to the formation of a surface layer exposing C-Br bonds that serve as initiators for the subsequent ATRP reaction. 

Lower surface densities of initiator were obtained by adding an intermediate step between the APTES and the BMPB treatments, which consists of immersing the APTES-coated substrates in a solution of PB (0.25 mL) and TEA (1.2 mL) in DCM (25 mL) for a prescribed time, rinsing in DCM/ethanol/water followed by drying. This resulted in the passivation of a fraction of the amine groups with the methyl-terminated, non ATRP active, PB molecule. The remaining -NH$_2$ groups were then functionalized as above. In the present work, we have used two different immersion times for this intermediate step, namely 1.5 and 6 minutes, in order to obtain brushes displaying two significantly lower densities as compared to the maximum density. 

\subsubsection*{Brush growth}
\label{subsubsec:growth}

PNIPAM brushes were grown from the initiator-grafted substrates using the recently established ARGET (activators regenerated by electron transfer) ATRP technique\cite{ARGET1,ARGET2}, that allows performing well-controlled radical polymerization reactions in ambient air. 

A solution of NIPAM (1.0 g) in ultrapure water (20mL) was prepared in a beaker containing a small stir bar. CuBr$_2$ (3mg) and the complexing ligand PMDETA (40 $\mu$L) were sequentially added while stirring. The solution turned blue at this stage. Upon addition of ascorbic acid (30mg), the solution became colorless, due to the reduction of Cu (II) into Cu (I). The solution was immediately poured onto the initiator-grafted substrates placed into petri dishes that were then closed during the polymerization reaction. Polymerization was left to proceed for a prescribed amount of time (typically between 2 minutes and 2 hours), after which the substrates were taken out of the reaction solution and thoroughly rinsed with water.

\subsection*{Comparison with other models of density profile}

\begin{figure}[h!]
\includegraphics[width=16cm]{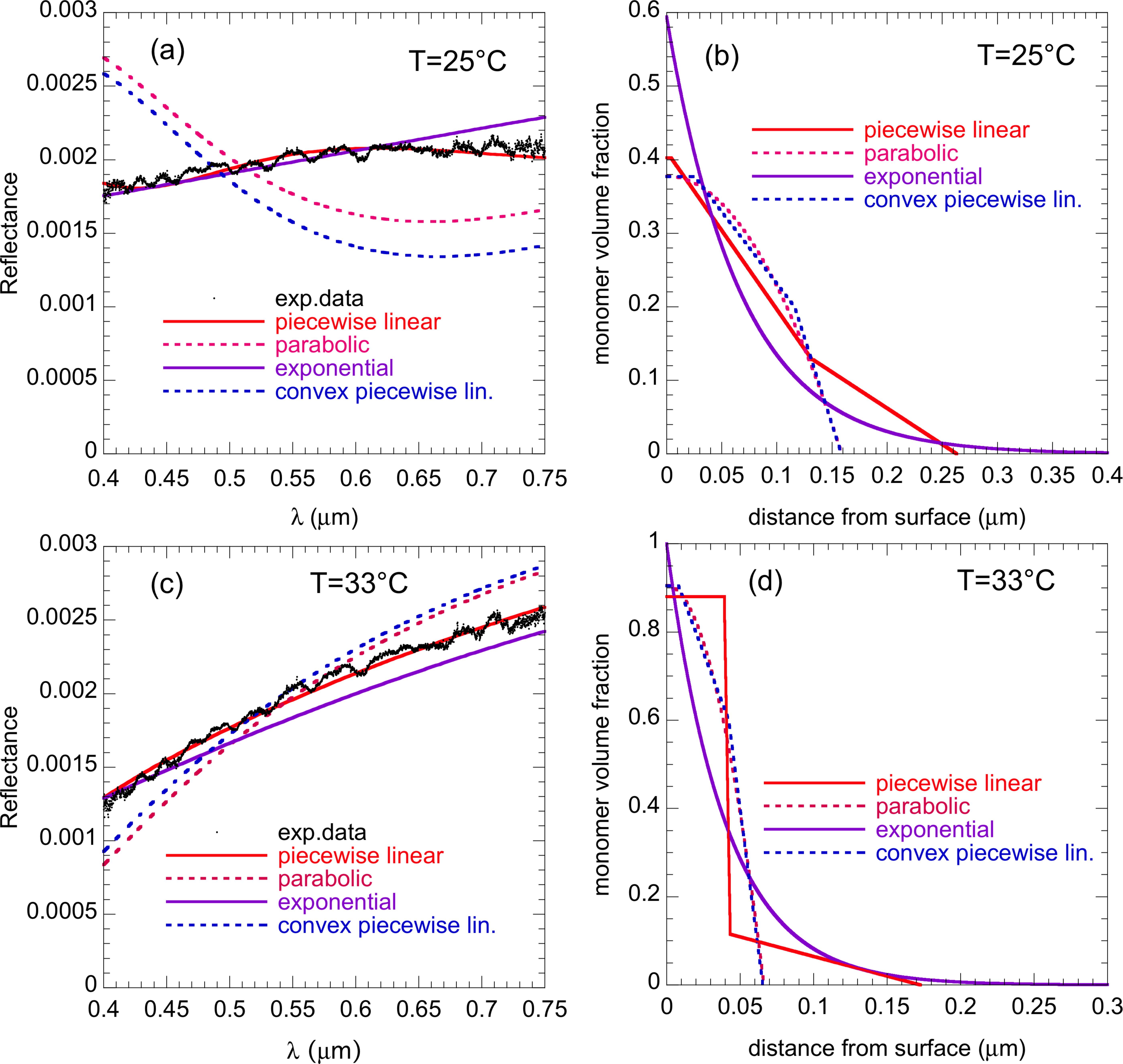}
\caption{\label{fig:otherprofiles} Experimental spectra obtained for the MD brush at 25$^{\circ}$C (a) and 33$^{\circ}$C (c), along with the best-fit spectra computed, as indicated on the panels, from an exponentially decaying profile ($\phi(z)=\phi_0e^{-z/h}$), a parabolic profile ($\phi(z)=\phi_0[1-(z/h)^2]$), a piecewise linear approximation of the parabola (convex piecewise lin.) and our piecewise linear model. The rightmost panels show such profiles at 25$^{\circ}$C (b) and 33$^{\circ}$C (d). All profiles satisfy the conservation of monomer amount ($\int_0^{+\infty}{\phi(z)dz=h_{dry}}$).}
\end{figure}

For the sake of completeness, we have performed a comparison of the data fits obtained with a parabolic profile (as theoretically expected for monodisperse brushes), a parabola approximated with our piecewise linear model, an exponentially decaying profile, and the best fit concave-shaped profile obtained from our model. 
Such a comparison is illustrated in Fig. \ref{fig:otherprofiles}, which shows that the first three profiles fail to reproduce the experimental data, while the latter satisfactorily fits the data over the whole spectral range. The results provided on Fig. \ref{fig:otherprofiles} also highlight the fact that the concavity of the density profile has a large influence on the fit quality (concave shapes fit better than convex ones). Finally, it can be seen that spectra computed with a parabolic density decay and with its approximation using our piecewise linear model exhibit highly similar features. This provides good support to the fact that the density profiles obtained with our simplified model are quantitatively reliable approximations of the actual density decays of the brushes.

\subsection*{Resolution limit}

\begin{figure}[h!]
\includegraphics[width=17cm]{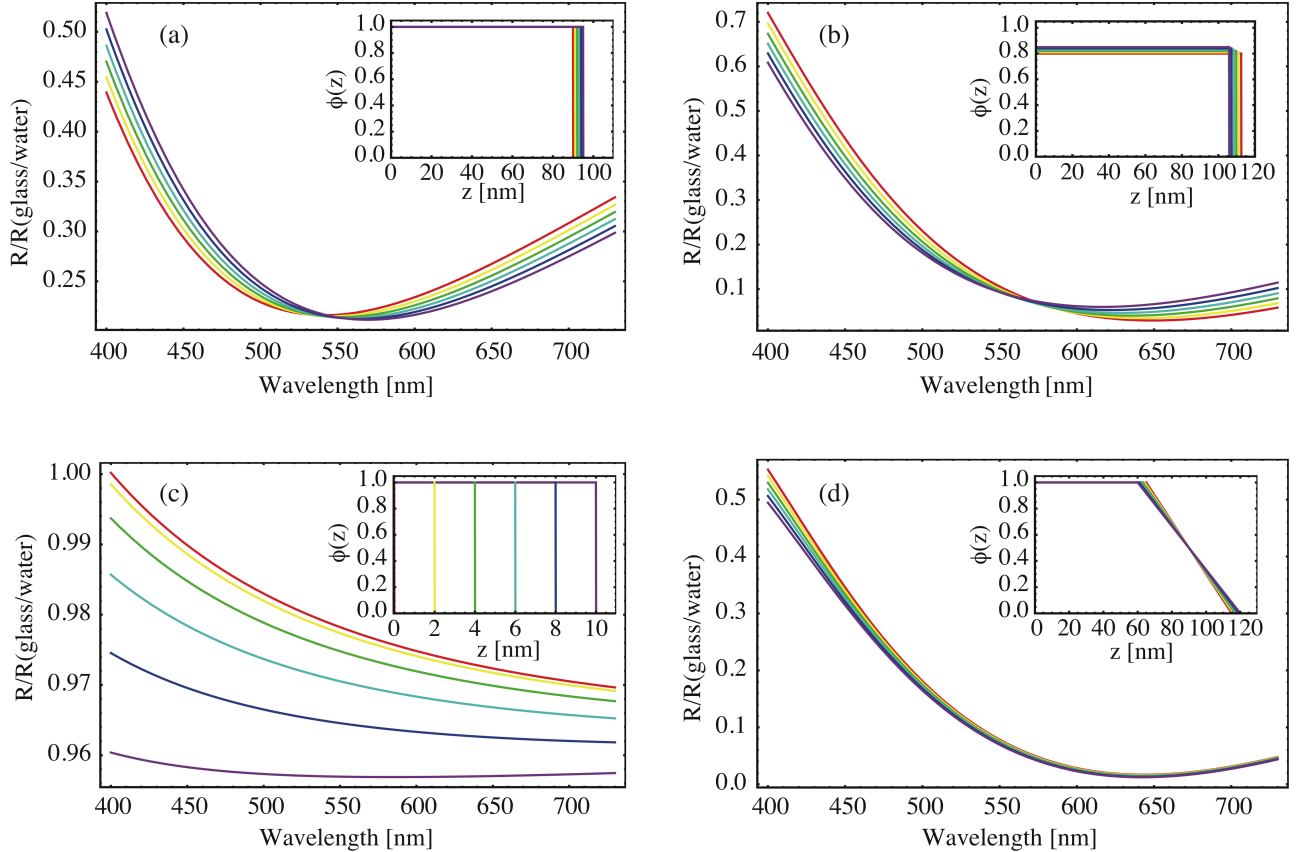}
\caption{\label{fig:ExamplesResolution} Examples of variations of the spectral reflectance curves with the brush parameters. The profiles corresponding to the computed spectra are shown in insets. Influence of the brush thickness: (a), in the 90-95nm range (1nm increment between each curve) and (c), in the 0-10nm range (2nm increment between each curve); (b), influence of brush density for a constant total amount of monomer in the 0.8-0.85 range (0.01 increment between each curve); (d), influence of brush density gradient for a decay length in the range 50-60nm (2nm increment between each curve).}
\end{figure}

Because spectral reflectance measurement is a coherent technique, resolution (or sensitivity to small changes in the profile) cannot be defined absolutely. An example is shown on Fig.~\ref{fig:ExamplesResolution}: spectral reflectance is very sensitive to small changes in thickness of the film (Fig.~\ref{fig:ExamplesResolution}(a)), or to its density (Fig.~\ref{fig:ExamplesResolution}(b)), when the film is a few tens of nanometer thick; very thin films ($<$5nm), on the other hand, are challenging to detect (Fig.~\ref{fig:ExamplesResolution}(c)). As a rule of thumb, for films thicker than 10nm, variations in thickness down to 1nm and changes in density down to 0.01 can be detected. Spectra are less sensitive to the slope of the decreasing intensity region, as illustrated in Fig.~\ref{fig:ExamplesResolution}(d): indeed sensitvity to the refractive index gradient (which modifies the reflectance amplitude of a given region) is less than that to the cumulated phase inbetween two interfaces (which is affected by the brush density and thickness).

\begin{figure}[h!]
\includegraphics[width=7cm]{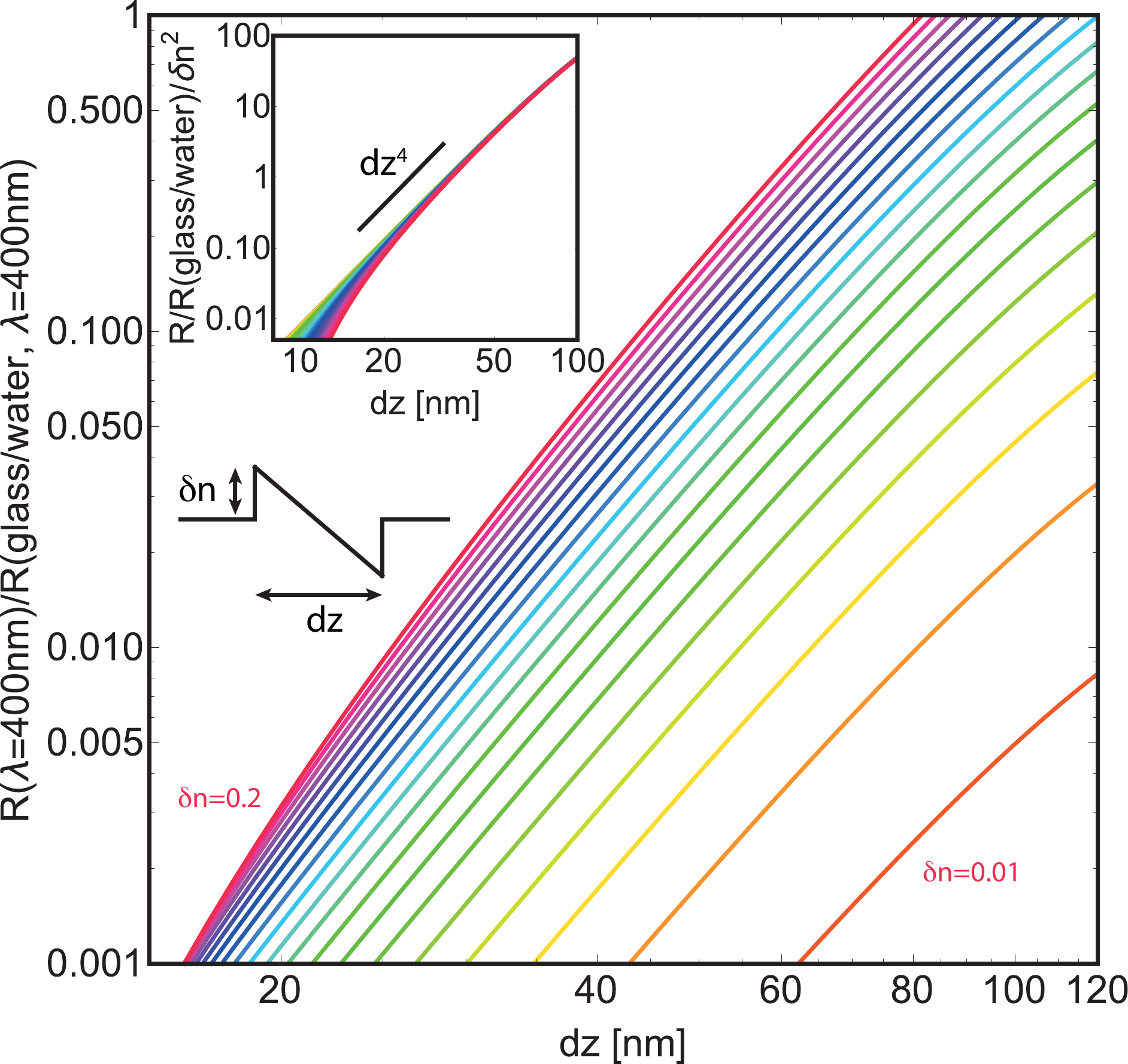}
\caption{\label{fig:Rwavelets} Resolution of the spectral reflectance measurements. The reflectance induced by a refractive index wavelet (shown as inset) is plotted as a function of its spatial extent and amplitude. The curves collapse when divided by the squared amplitude (inset).}
\end{figure}

In order to assess in a more general manner the resolution that can be achieved for the reconstruction of the phase profile, we have investigated the sensitivity of this method to ``wavelets" of monomer density, as represented on Fig.~\ref{fig:Rwavelets}: indeed, when added to a density profile, such density modulation does not induce any phase difference on the rest of the profile, so that its contribution to the signal can be better isolated and characterized. Furthermore, the difference between a one-phase and a two-phase density profile can roughly be expressed as a wavelet of this type.

This test density profile can be described fully by two parameters : the amplitude of the refractive index variation, $\delta n$, and its extent, $dz$. We have plotted on Fig.~\ref{fig:Rwavelets} the reflectance of such profile as a function of these two parameters at $\lambda=400$nm: the reflectance roughly scales as $\delta n^2 \times dz^4$, and if we set the detection limit to about $0.1\%$ of the reflectivity of a glass/water interface, the minimal detected size of a wavelet of amplitude $\delta n=0.05$ (or equivalently $\delta \Phi=0.4$) is $dz\approx 30$nm. This is consistent with the fact that for brushes of thickness smaller than 60nm, we can equivalently fit the measured reflectance spectra with a one-phase and a two-phase profile.

\subsection*{Phase separation around the LCST}

\begin{figure}[h!]
\includegraphics[width=8cm]{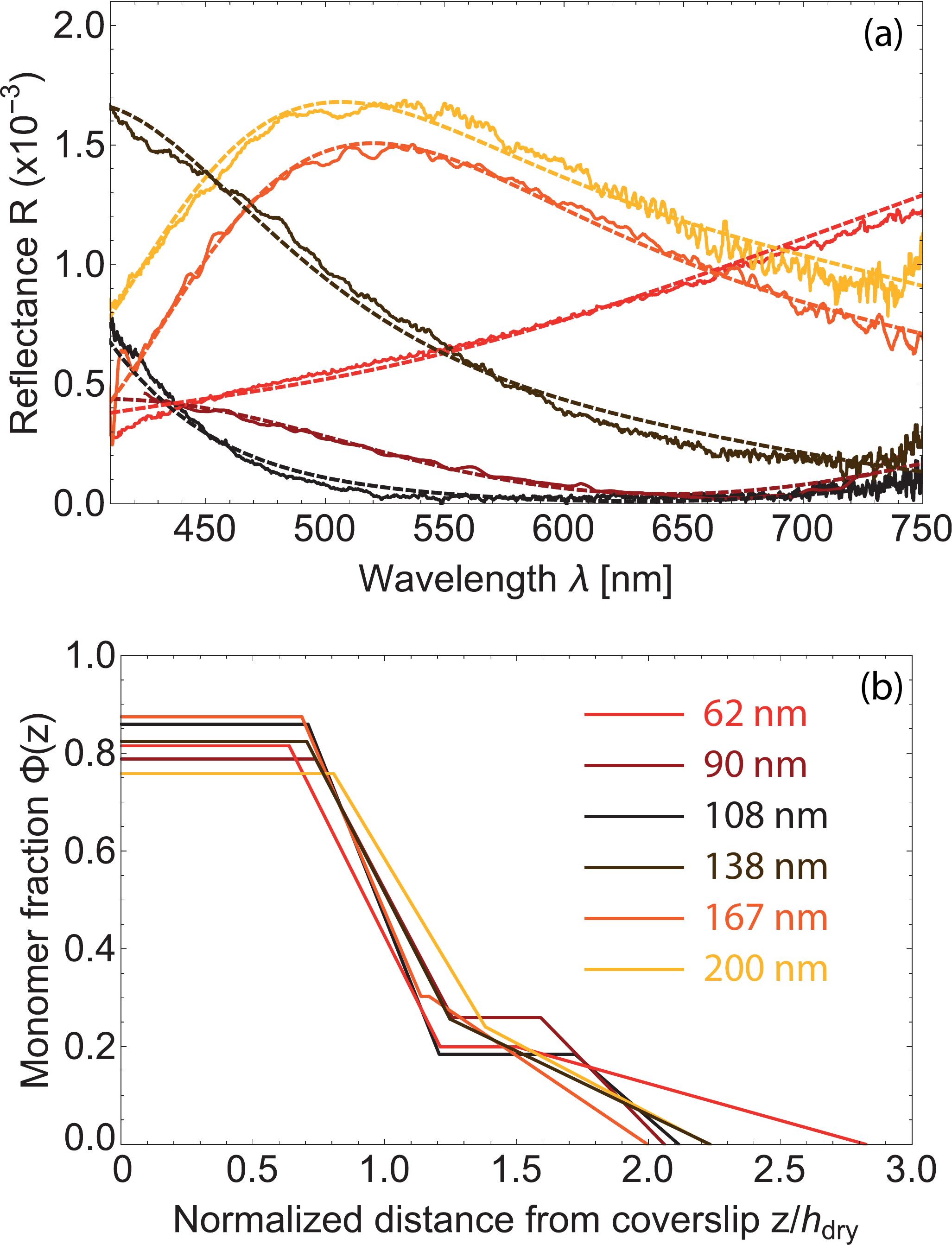}
\caption{\label{fig:fitsatthetransition} (a), reflectance spectra of brushes HD1, HD2, HD3, HD4, HD5 and HD7 at 32$^{\circ}$C (solid lines) and fits (dashed lines) with the two-phase profiles plotted in (b), showing a consistent phase separation for all six thicknesses. Brushes of smaller dry thicknesses do not appear on the figure as their potential phase separation cannot be resolved using our setup.}
\end{figure}

On Fig.~\ref{fig:fitsatthetransition} we have plotted the spectral reflectance and corresponding fits of 6 dense polymer brushes around the LCST: despite minor variations between the different profiles, the phase separation is consistently observed. 

\end{document}